\documentclass[final,3p,times,twocolumn]{elsarticle}

\usepackage{amssymb}
\usepackage{amsthm,bm}
\usepackage{graphicx,psfrag}
\usepackage[rightcaption]{sidecap}
\usepackage[percent]{overpic}

\journal{Journal of Computational Physics}

\begin{document}

\begin{frontmatter}


\title{Regularized characteristic boundary conditions for the Lattice-Boltzmann methods at high Reynolds number flows}

\author[ISAE,Altran,Cerfacs]{Gauthier Wissocq}
\author[ISAE]{Nicolas Gourdain}
\author[UGEN]{Orestis Malaspinas}
\author[Altran]{Alexandre Eyssartier}

\address[ISAE]{ISAE, Dpt. of Aerodynamics, Energetics and Propulsion, Toulouse, France}
\address[Altran]{Altran, DO ME, Blagnac, France}
\address[Cerfacs]{Centre Europ\'{e}en de Recherche et de Formation Avanc\'{e}e en Calcul Scientifique (CERFACS), CFD Team, 42 avenue Gaspard Coriolis, 31057 Toulouse Cedex 01, France}
\address[UGEN]{SPC - Centre Universitaire d'Informatique, Universit{\'e} de Gen{\`e}ve 7, route de Drize, CH-1227 Switzerland}

\begin{abstract}
This paper reports the investigations done to adapt the Characteristic Boundary Conditions (CBC) to the Lattice-Boltzmann formalism for high Reynolds number applications. Three CBC formalisms are implemented and tested in an open source LBM code: the baseline local one-dimension inviscid (BL-LODI) approach, its extension including the effects of the transverse terms (CBC-2D) and a local streamline approach in which the problem is reformulated in the incident wave framework (LS-LODI). Then all implementations of the CBC methods are tested for a variety of test cases, ranging from canonical problems (such as 2D plane and spherical waves and 2D vortices) to a 2D NACA profile at high Reynolds number ($Re = 10^5$), representative of aeronautic applications. The LS-LODI approach provides the best results for pure acoustics waves (plane and spherical waves). However, it is not well suited to the outflow of a convected vortex for which the CBC-2D associated with a relaxation on density and transverse waves provides the best results. As regards numerical stability, a regularized adaptation is necessary to simulate high Reynolds number flows. The so-called regularized FD (Finite Difference) adaptation, a modified regularized approach where the off-equilibrium part of the stress tensor is computed thanks to a finite difference scheme, is the only tested adaptation that can handle the high Reynolds computation.
\end{abstract}


\end{frontmatter}

\section*{Introduction}
A better understanding of turbulent unsteady flows is a necessary step towards a breakthrough in the design of modern aircraft and propulsive systems. Due to the difficulty of predicting turbulence with complex geometry, the flow that develops in these engines remains difficult to predict. At this time, the most popular method to model the effect of turbulence is still the Reynolds Averaged Navier-Stokes (RANS) approach. However there is some evidence that this formalism is not accurate enough, especially when a description of time-dependent turbulent flows is desired (high incidence angle, laminar-to-turbulent transition, etc.)~\cite{Tucker:2014}. With the increase in computing power, Large Eddy Simulation (LES) applied to the Navier-Stokes equations emerges as a promising technique to improve both knowledge of complex physics and reliability of flow solver predictions~\cite{Tucker:2014}. It is still the most popular and mature approach to describe the behavior of turbulent flow in complex geometries ({\it e.g.} aircraft and gas turbines). However, the resolution of the NS equations requires to add artificial dissipation to ensure numerical stability~\cite{Tucker:2014}. The consequence is an over-dissipation which affects the flow and limits the capability to transport flow patterns (like turbulence) on a long distance. In some specific cases, like aero-acoustic (far-field noise), NS can thus face some difficulties to predict the flow.

In this context, there is an increasing interest in the fluid dynamics community for emerging methods, based on the Lattice Boltzmann approach~\cite{Chen_AnnuRevFluid_30_1998, Succi_2001, Lallemand_PhysRevE_61_2000}. The Lattice-Boltzmann Method (LBM) has already demonstrated its potential for complex geometries, thanks to immersed boundary conditions (that allow the use of cartesian grids) and low dissipation properties required for capturing the small acoustic pressure fluctuations \cite{Buick_EPL_43_1998, Marie_JCP_228_2009}. LBM also provides the advantage of an easy parallelization, making it well suited for High-Performance Computing \cite{Heuveline_CMAP_58_2009}. However, the most widely used Lattice-Boltzmann models still suffer from weaknesses like a lack of robustness for high Mach number flows ($M~>~0.4$), a limitation to low compressible isothermal flows \cite{Chen_AnnuRevFluid_30_1998} and the use of artificial boundary conditions (Dirichlet/Neumann types can lead to the reflection of outgoing acoustic waves that have a significant influence on the flow field \cite{Colonius_AnnuRevFluid_36_2004, Bodony_JCP_212_2006, Israeli_JCP_41_1981}). While the use of artificial boundary conditions is also critical for NS methods, it is more problematic for LBM due to the low dissipation of the method.

A potential way to avoid unphysical acoustic reflections at the boundary is to use a "sponge layer" inside the computational domain, on which artificial dissipation (by upwinding) is introduced or physical viscosity is increased (viscosity sponge zones). Acoustic waves (physical or not) are thus damped in such a zone, which allows to eliminate or limit numerical reflections~\cite{Bodony_JCP_212_2006}. This solution has however important drawbacks. First the calibration of sponge layers is difficult, as a balance must be found between a brutal increase of the viscosity (that will generate acoustic reflections) and a too low dissipation that will not be effective. Such sponge layers also have an impact on the computational cost, since a part of the domain is dedicated to slowly increasing the viscosity. Last, some boundary conditions cannot be treated with a sponge layer, for instance an inlet with turbulence injection.

For NS methods, a successful approach is the use of non-reflective boundary conditions based on a treatment of the characteristic waves of the local flow \cite{Poinsot_JCP_101_1992, Yoo_CTM_9_2005, Lodato_JCP_227_2008}. However, the extension of this approach to LBM is not straightforward given the difficulty to find a bridge between the LBM that describes the world at the mesoscopic level (a population of particles) and the NS world based on a macroscopic description of the flow. But some progress has recently been made on the adaptation of characteristic boundary conditions to the LBM formalism. Izquierdo and Fueyo~\cite{Izquierdo_PhysRevE_78_2008} used a pressure antibounceback boundary condition~\cite{Ginzburg:2008} adapted to the multiple relaxation time (MRT) collision scheme~\cite{Dhumieres:1992} to impose the Dirichlet density and velocity conditions given by the local one-dimensional inviscid (LODI) equations, which provided non-reflective outflow boundary conditions for one-dimensional waves. More recently, Jung \textit{et al.}~\cite{Jung:2015} extended the previous work to include the effects of transverse and viscous terms in the Characteristic Boundary Conditions (CBC) and showed good performance for vortex outflow. Meanwhile, Heubes \textit{et al.}~\cite{Heubes_JCAM_262_2014} adapted the solution given by a modified-Thompson approach by imposing the corresponding equilibrium populations and Schlaffer~\cite{Schlaffer:2013} assessed a modified Zou/He boundary condition~\cite{Zou_PhysFluids_9_1997}. 

Still, previous researches are limited to low Reynolds number applications while Latt \textit{et al.} showed that increasing the Reynolds number can have drastic impact on the numerical stability of the LBM boundary condition~\cite{Latt_PhysRevE_77_2008}. Even when the MRT collision is used, as in \cite{Izquierdo_PhysRevE_78_2008, Ginzburg:2008}, the numerical stability of the characteristic boundary conditions has not been demonstrated. The aim of this study is thus to develop a numerically stable adaptation of the CBC to the LBM formalism taking advantage of the regularized collision scheme~\cite{Latt:2006}, which has proved to be numerically stable at high Reynolds number flows~\cite{Malaspinas:2015}, and the corresponding regularized boundary conditions~\cite{Latt_PhysRevE_77_2008}. Different kinds of CBC will also be evaluated. This article is structured as follows. The first section describes the LBM framework. Then, the second section presents three kinds of CBCs and three possible adaptations to the LBM formalism for 2D problems in the low compressible isothermal case. In the third section, these models are assessed for simple cases: normal, oblique and spherical waves and a convected vortex at $Re=10^3$. Finally, the method is assessed on a high Reynolds number application: a NACA0015 airfoil at $Re=10^5$.

\section{Numerical method and governing equations}

\subsection*{Lattice Boltzmann framework for isothermal flows}
A description of the Lattice-Boltzmann Method can be found in \cite{Chen_AnnuRevFluid_30_1998, Succi_2001, Lallemand_PhysRevE_61_2000}. The governing equations describe the evolution of the probability density of finding a set of particles with a given microscopic velocity at a given location: 

\begin{equation}
\label{LBM equation}
f_i(\mathbf{x}+\mathbf{c_i} \Delta t , t+\Delta t) = f_i(\mathbf{x},t) + \Omega_{i} (\mathbf{x},t)
\end{equation}
for $0\leq i < q$, where $\mathbf{c_i}$ is a discrete set of $q$ velocities, $f_i(\mathbf{x},t)$ is the discrete single particle distribution function corresponding to $\mathbf{c_i}$ and $\Omega_{i}$ is an operator representing the internal collisions of pairs of particles.

Macroscopic values such as density, $\rho$, and the flow velocity, $\mathbf{u}$, can be deduced from the set of probability density functions $f_i(\mathbf{x},t)$, such as:
\begin{equation}
\rho = \sum_{i=0}^{q-1} f_i, \ \ \ \ \ \rho \mathbf{u} =\sum_{i=0}^{q-1} f_i \mathbf{c_i}.
\end{equation}

\begin{center}
\begin{figure}[h!]
\includegraphics[width=0.49\textwidth]{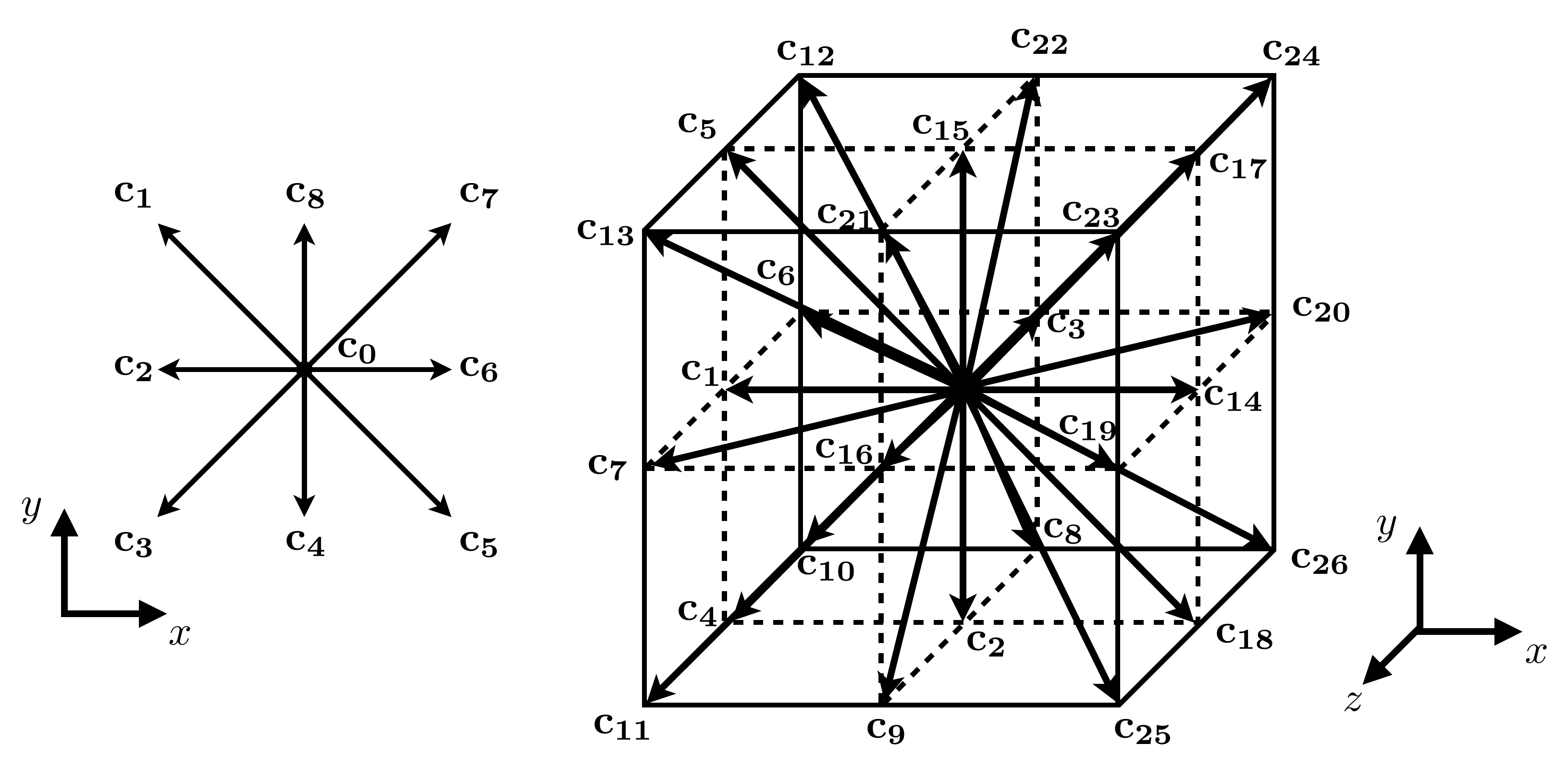}
\caption{Schematic plot of velocity directions of the D2Q9 model (left) and the D3Q27 model (right)\label{fig:D2Q9D3Q27}}
\end{figure}
\end{center}

Some of the most popular choices for the set of velocities are D2Q9 and D3Q27 lattices, respectively 9 velocities in 2D and 27 velocities in 3D (see Fig.~\ref{fig:D2Q9D3Q27}). For both of these lattices, the sound speed in lattice units (normalized by the ratio between the spatial resolution and the time step $\Delta x/\Delta t$) is given by $c_s = 1/\sqrt{3}$ \cite{Succi_2001}.

The collision operator $\Omega_{i}$ is usually modelled with the Bhatnagar-Gross-Krook (BGK) approximation \cite{Bhatnaghar_PhysRev_94_1954}, which consists in a relaxation, with a relaxation time $\tau$, of every population to the corresponding equilibrium probability density function $f_i^{(eq)}$:
\begin{equation}
\Omega_{i} = -\frac{1}{\tau}\left[f_i(x,t)-f_i^{(eq)}(x,t)\right].
\end{equation}

The equilibrium distribution function $f_i^{(eq)}$ is a local function that only depends on density and velocity in the isothermal case. It can be computed thanks to a second order development of the Maxwell-Boltzmann equilibrium function \cite{Qian_EPL_17_1992}:
\begin{equation}
f_i^{(eq)} = w_i \rho \left[ 1+\frac{\mathbf{c_i}\cdot \mathbf{u}}{c_s^2} + \left( \frac{\mathbf{c_i} \cdot \mathbf{u}}{2c_s^2} \right)^2- \frac{\mathbf{u}^2}{2c_s^2} \right],
\end{equation}
where $w_i$ are the gaussian weights of the lattice.

A Chapman-Enskog expansion, based on the assumption that $f_i$ is given by the sum of the equilibrium distribution plus a small perturbation $f^{(1)}_i$
\begin{equation}
 f_i=f_i^{(eq)}+f_i^{(1)},\hbox{ with } f_i^{(1)}\ll f_i^{(eq)} \label{eq_fi_ce},
\end{equation}
can be applied to (\ref{LBM equation}) in order to recover the exact Navier-Stokes equation for quasi-incompressible flows in the limit of long-wavelength \cite{Chapman_1952}. The pressure is thus given by $p=c_s^2 \rho$ and the kinematic viscosity is linked to the BGK relaxation parameter through
\begin{equation}
\nu = c_s^2(\tau - \frac{1}{2}).
\end{equation}
The Chapman--Enskog expansion also relates the second order tensor $\bm{\Pi^{(1)}}$ defined as
\begin{equation}
\bm{\Pi^{(1)}}=\sum_i\mathbf{c_i}\mathbf{c_i}f^{(1)}_i,\label{eq_pi1_f1}
\end{equation}
with the strain rate tensor $\mathbf{S}=(\bm{\nabla}\mathbf{u}+(\bm{\nabla}\mathbf{u})^T)/2$ through the relation
\begin{equation}
\bm{\Pi^{(1)}}=-2 c_s^2 \rho\tau \mathbf{S}. \label{pi1_grad}
\end{equation}
In turn to the leading order $f_i^{(1)}$ can be approximated by
\begin{equation}
 f_i^{(1)}\cong \frac{w_i}{2 c_s^2}\mathbf{Q}_i:\bm{\Pi^{(1)}},\label{eq_f1_pi1}
\end{equation}
where $\mathbf{Q}_i\equiv\left(\mathbf{c}_i\mathbf{c}_i-c_s^2\mathbf{I}\right)$. The colon symbol stands for the double contraction operator and $\mathbf{I}$ is the identity matrix. \newline
A regularization step consisting in the reconstruction of the off-equilibrium parts using (\ref{eq_f1_pi1}) and (\ref{eq_pi1_f1}) can improve the precision and the numerical stability of the single relaxation time BGK collision~\cite{Latt:2006, Malaspinas:2015}. This model will be used in the next parts for high Reynolds number flow simulations.

\subsection*{The Palabos open-source library}
The LBM flow solver used in this work is the Palabos\footnote{Copyright 2011-2012 FlowKit Ltd.} open-source library. The Palabos library is a framework for general-purpose CFD with a kernel based on the lattice Boltzmann method. The use of C++ code makes it easy to install and to run on every machine. It is thus possible to set up fluid flow simulations with relative ease and to extend the open-source library with new methods and models, which is of paramount importance for the implementation of new boundary conditions. The numerical scheme is divided in two steps:
\begin{itemize}
\item A collision step where the BGK model is applied:
\begin{equation}
f_i(\mathbf{x}, t+\frac{1}{2}) = f_i(\mathbf{x},t) + \frac{1}{\tau}\left[f_i^{(eq)}(\mathbf{x},t)-f_i(\mathbf{x},t)\right],\ 
\end{equation}
with $f_i^{(eq)}$ computed using the macroscopic values at time $t$ and $f_i$ can be regularized in order to increase numerical stability for high Reynolds number flows.
\item A streaming step:
\begin{equation}
f_i(\mathbf{x}+\mathbf{c_i}, t+1) = f_i(\mathbf{x}, t+\frac{1}{2}).
\end{equation}
\end{itemize}

The streaming step consists in an advection of each discrete population to the neighbor node located in the direction of the corresponding discrete velocity. Since a boundary node has less neighbors than an internal node (less than 9 neighbors in 2D or 27 neighbors in 3D), some populations are missing at the boundary after each iteration. These populations need to be reconstructed, which is the purpose of the implementation of boundary conditions in LBM. Up to now, different methods can be used in Palabos, such as regularized BC \cite{Malaspinas_CompFluids_49_2011}
 or Zou/He BC \cite{Zou_PhysFluids_9_1997}
to implement open boundaries. However, none of them can be used as they stand for an outflow boundary condition and the use of sponge zones is necessary to avoid non physical reflections. The next sections will aim at developing a more natural boundary condition that minimize acoustic reflections for an outflow boundary type, based on the Characteristic Boundary Conditions (CBC).

\section{Adaptation of Characteristic Boundary Conditions to the LBM formalism}

One of the most popular methods in the NS community for subsonic non reflective outflow boundary conditions is the CBC method \cite{Poinsot_JCP_101_1992}. The adaptation of the CBC to the LBM formalism is presented here for an isothermal flow, in lattice units (normalized by $\Delta x$ and $\Delta t$). Acoustic waves thus propagate at the constant lattice sound speed $c_s$. In the isothermal case, pressure is defined as $p=c_s^2 \rho$. Three CBC methods will be introduced below: the local one-dimensional inviscid (LODI) approximation, a 2D extension of the LODI approximation including transverse waves and a last method called local-streamline LODI.\newline

\subsection*{LODI Approximation (Baseline LODI)}

\begin{center}
\begin{figure}[h]
\includegraphics[width=0.49\textwidth]{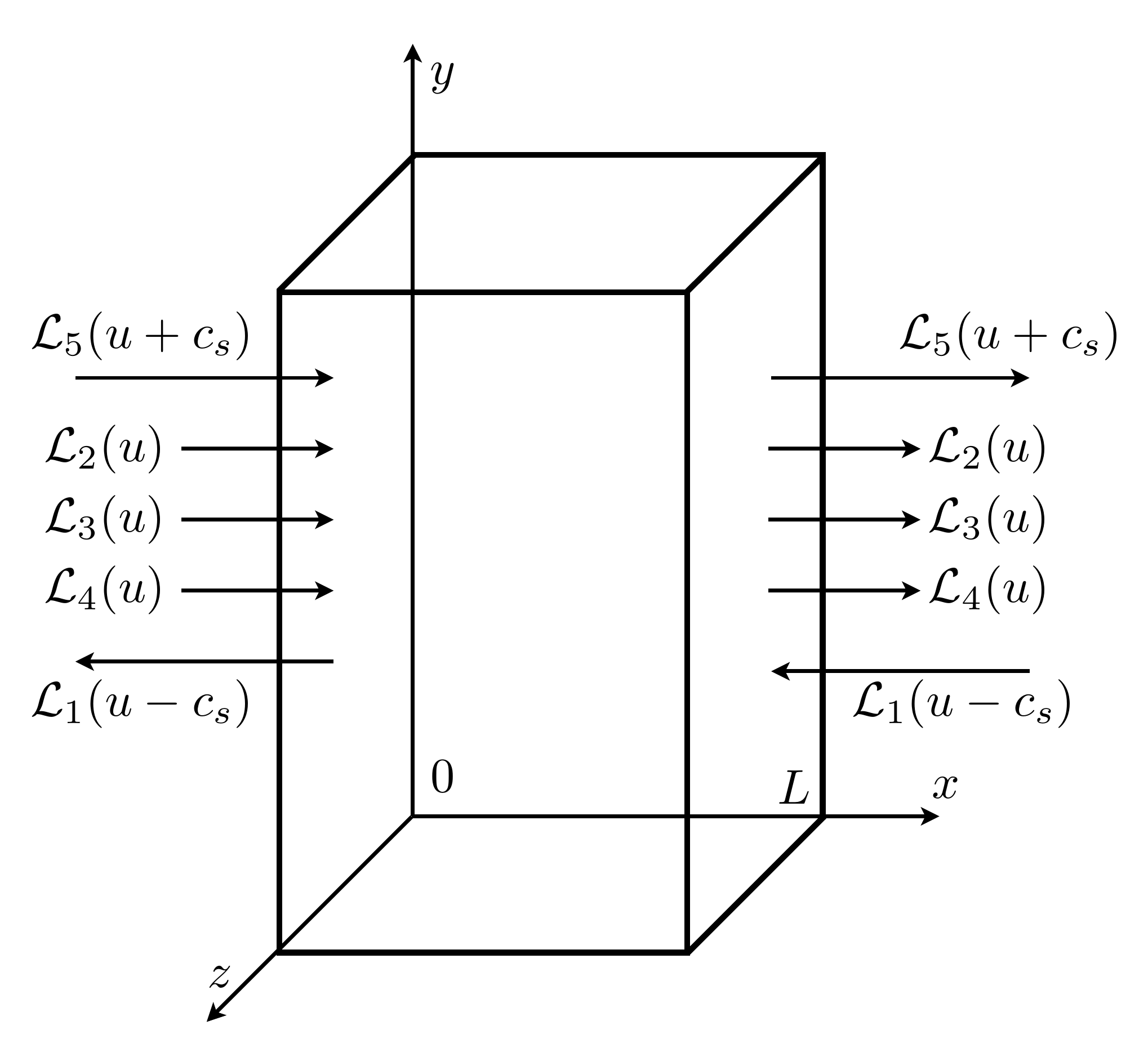}
\caption{Waves leaving and entering the computational domain through an inlet plane ($x=0$) and an outlet plane ($x=L$) for a subsonic flow \cite{Poinsot_JCP_101_1992} in lattice units.\label{fig:CBC}}
\end{figure}
\end{center}

Let us consider a domain outlet located at $x=L$, as descripted on Fig.~\ref{fig:CBC}. A diagonalisation of the x-derivative terms in the Navier-Stokes equation allows to define five waves $\mathcal{L}_i$ that propagate respectively at velocity $u-c_s$, $u$, $u$, $u$ and $u+c_s$, where $u$ is the x-component (streamwise) of the non-dimensional macroscopic velocity $\mathbf{u}= [u, v, w]$. These waves are represented on Fig.~\ref{fig:CBC} on the inlet ($x=0$) and outlet ($x=L$) of a computational domain.

At the outlet ($x=L$ on Fig.~\ref{fig:CBC}), $\mathcal{L}_2$, $\mathcal{L}_3$, $\mathcal{L}_4$ and $\mathcal{L}_5$ leave the computational domain and are obtained with the general expression of characteristic waves:
\begin{equation}
\mathcal{L}_2=u\left( c_s^2 \frac{\partial \rho}{\partial x} - \frac{\partial p}{\partial x} \right)=0, 
\end{equation}
\begin{equation}
\mathcal{L}_3=u \frac{\partial v}{\partial x}, 
\end{equation}
\begin{equation}
\mathcal{L}_4=u \frac{\partial w}{\partial x}, 
\end{equation}
\begin{equation}
\mathcal{L}_5=(u+c) \left( \frac{\partial p}{\partial x} + \rho c_s \frac{\partial u}{\partial x} \right).
\end{equation}
Let us notice that $\mathcal{L}_2$, the entropy wave, is null in the isothermal case.The $x$-derivative terms can be computed using the interior points by one-sided finite difference. The treatment of $\mathcal{L}_1$ is different : since it comes from the outside, it can not be computed using the interior points. The perfectly non-reflecting case is obtained by fixing $\mathcal{L}_1=0$, which ensures eliminating the incoming wave. However, this is known to be unstable because of lack of control of the outlet flow variables. A simple way to ensure well-posedness is to set
\begin{equation}
\label{L1}
\mathcal{L}_1=K_1(p-p_{\infty}), 
\end{equation}
where $K_1=\sigma (1-M^2)c_s/L$, $p_\infty$ is the target pressure at the outlet, $\sigma$ is a constant, $M$ is the maximum Mach number in the flow and $L$ is a characteristic size of the domain~\cite{Poinsot_JCP_101_1992}.\newline

The time-derivative of the primitive variables can be computed in function of the wave amplitudes by examining a LODI problem :
\begin{equation}
\label{LODIrho}
\frac{\partial \rho}{\partial t} + \frac{1}{c_s^2} \left[ \mathcal{L}_2 + \frac{1}{2}(\mathcal{L}_5+\mathcal{L}_1) \right]=0, 
\end{equation}
\begin{equation}
\label{LODIp}
\frac{\partial p}{\partial t} + \frac{1}{2}(\mathcal{L}_5+\mathcal{L}_1)=0, 
\end{equation}
\begin{equation}
\label{LODIu}
\frac{\partial u}{\partial t}+\frac{1}{2\rho c_s}(\mathcal{L}_5-\mathcal{L}_1)=0, 
\end{equation}
\begin{equation}
\label{LODIv}
\frac{\partial v}{\partial t} + \mathcal{L}_3 = 0, 
\end{equation}
\begin{equation}
\label{LODIw}
\frac{\partial w}{\partial t}+\mathcal{L}_4=0.
\end{equation}
In the isothermal case, (\ref{LODIrho}) and (\ref{LODIp}) are equivalent. Finally, with a temporal discretization using an explicit second-order scheme, the physical values that must be imposed at the next time step in order to avoid acoustic reflections can be computed. This implementation will be referred to as the baseline LODI (BL-LODI) in the rest of the paper.

\subsection*{LODI approximation including transverse terms}

The previous relations are perfectly non-reflecting for the only case of a pure 1D plane wave. For a non-normal wave, the LODI approximation is not verified and a reflected wave, all the more important as the incidence increases, can appear. To take this phenomenon into account, a possible solution is to add the influence of transverse waves in the LODI equations~\cite{Yoo:2007, Jung:2015}:
\begin{equation}
\label{LODITransverserho}
\frac{\partial \rho}{\partial t} + \frac{1}{2c_s^2}(\mathcal{L}_5+\mathcal{L}_1)=\frac{1}{2c_s^2}(\mathcal{T}_5+\mathcal{T}_1), 
\end{equation}
\begin{equation}
\label{LODITransverseu}
\frac{\partial u}{\partial t}+\frac{1}{2\rho c_s}(\mathcal{L}_5-\mathcal{L}_1)=\frac{1}{2\rho c_s}(\mathcal{T}_5-\mathcal{T}_1), 
\end{equation}
\begin{equation}
\label{LODITransversev}
\frac{\partial v}{\partial t} + \mathcal{L}_3 = \mathcal{T}_3, 
\end{equation}
\begin{equation}
\label{LODITransversew}
\frac{\partial w}{\partial t}+\mathcal{L}_4=\mathcal{T}_4.
\end{equation}
The transverse waves can be computed as follows:
\begin{equation}
\label{T1}
\mathcal{T}_1=-\left[ \mathbf{u_t} \cdot \boldsymbol{\nabla_t}p +p\boldsymbol{\nabla_t}\cdot \mathbf{u_t} -\rho c_s \mathbf{u_t} \cdot \boldsymbol{\nabla_t}u \right],
\end{equation}
\begin{equation}
\label{T3}
\mathcal{T}_3=-\left[ \mathbf{u_t}\cdot \boldsymbol{\nabla_t}v + \frac{1}{\rho}\frac{\partial p}{\partial y} \right],
\end{equation}
\begin{equation}
\label{T4}
\mathcal{T}_4=-\left[ \mathbf{u_t}\cdot \boldsymbol{\nabla_t}w + \frac{1}{\rho}\frac{\partial p}{\partial z} \right],
\end{equation}
\begin{equation}
\label{T5}
\mathcal{T}_5=-\left[ \mathbf{u_t} \cdot \boldsymbol{\nabla_t}p +p\boldsymbol{\nabla_t}\cdot \mathbf{u_t} +\rho c_s \mathbf{u_t} \cdot \boldsymbol{\nabla_t}u \right],
\end{equation}
where $\mathbf{u_t}=[v,w]$ and $\boldsymbol{\nabla_t}=[\partial_y, \partial_z]$.
The second transverse wave $\mathcal{T}_2$ is not introduced here since it is null in the isothermal case, as well as $\mathcal{L}_2$. The non reflective outflow boundary condition needs now to be set as:
\begin{equation}
\mathcal{L}_1=K_1(p-p_{\infty}) - K_2(\mathcal{T}_1-\mathcal{T}_{1,exact}) + \mathcal{T}_1, 
\end{equation}
with $K_1=\sigma (1-M^2)c_s/L$, $K_2$ should be equal to the Mach number of the mean flow and $\mathcal{T}_{1,exact}$ is a desired steady value of $\mathcal{T}_1$ ~\cite{Yoo:2007}. In the rest of the paper, this method will be named 2D-CBC in the perfectly non-reflecting case ($K_1=K_2=0$) and 2D-CBC relaxed when a relaxation is done on $\mathcal{L}_1$.

\subsection*{Local streamline LODI}

\begin{figure}[h!]
\indent (a) geometry-based frame $\mathcal{R}$\\
\put(30,0){\includegraphics[width=0.295\textwidth]{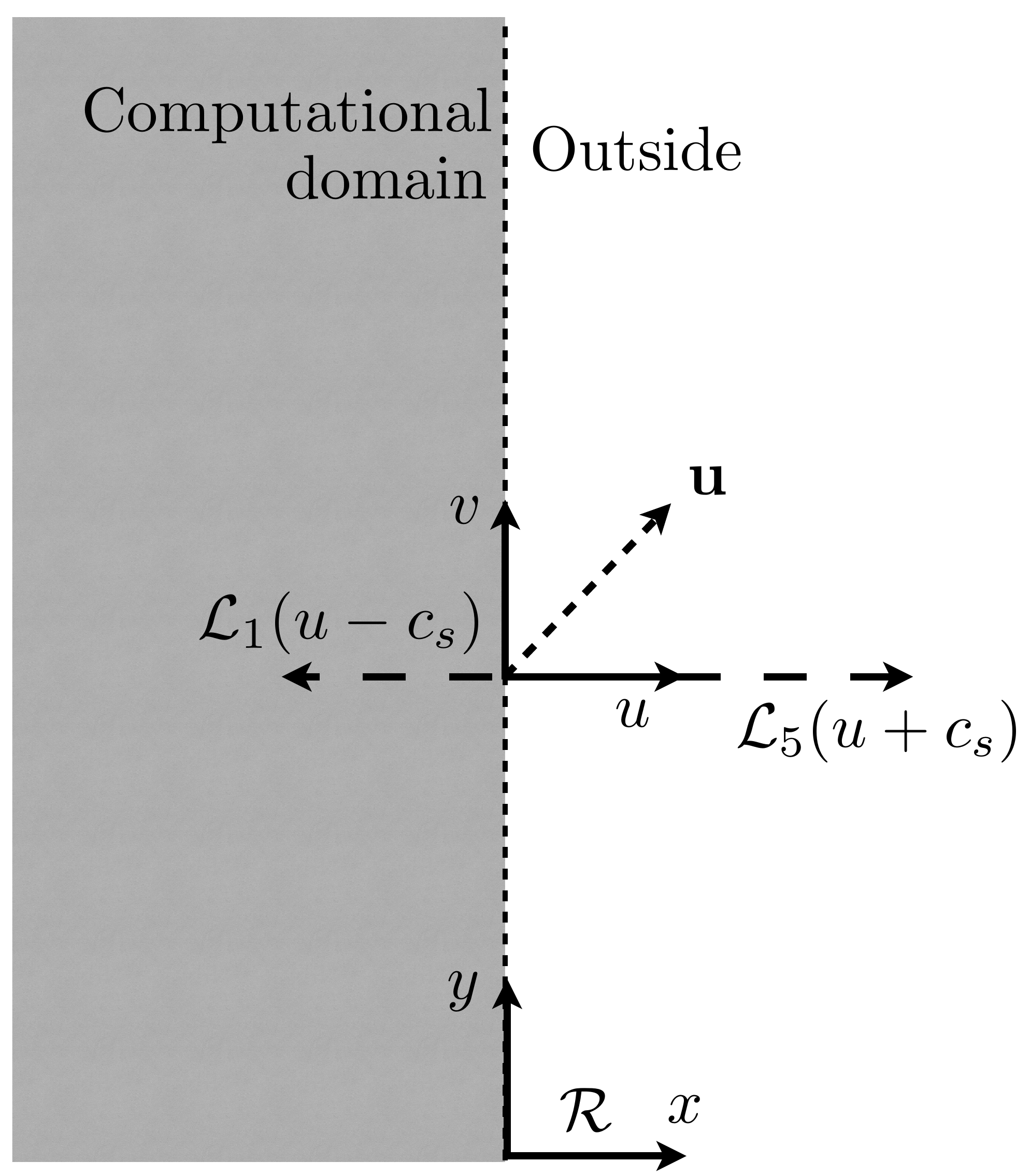}} \\
\indent (b) local streamline-based frame $\tilde \mathcal{R}$\\
\put(30,0){\includegraphics[width=0.27\textwidth]{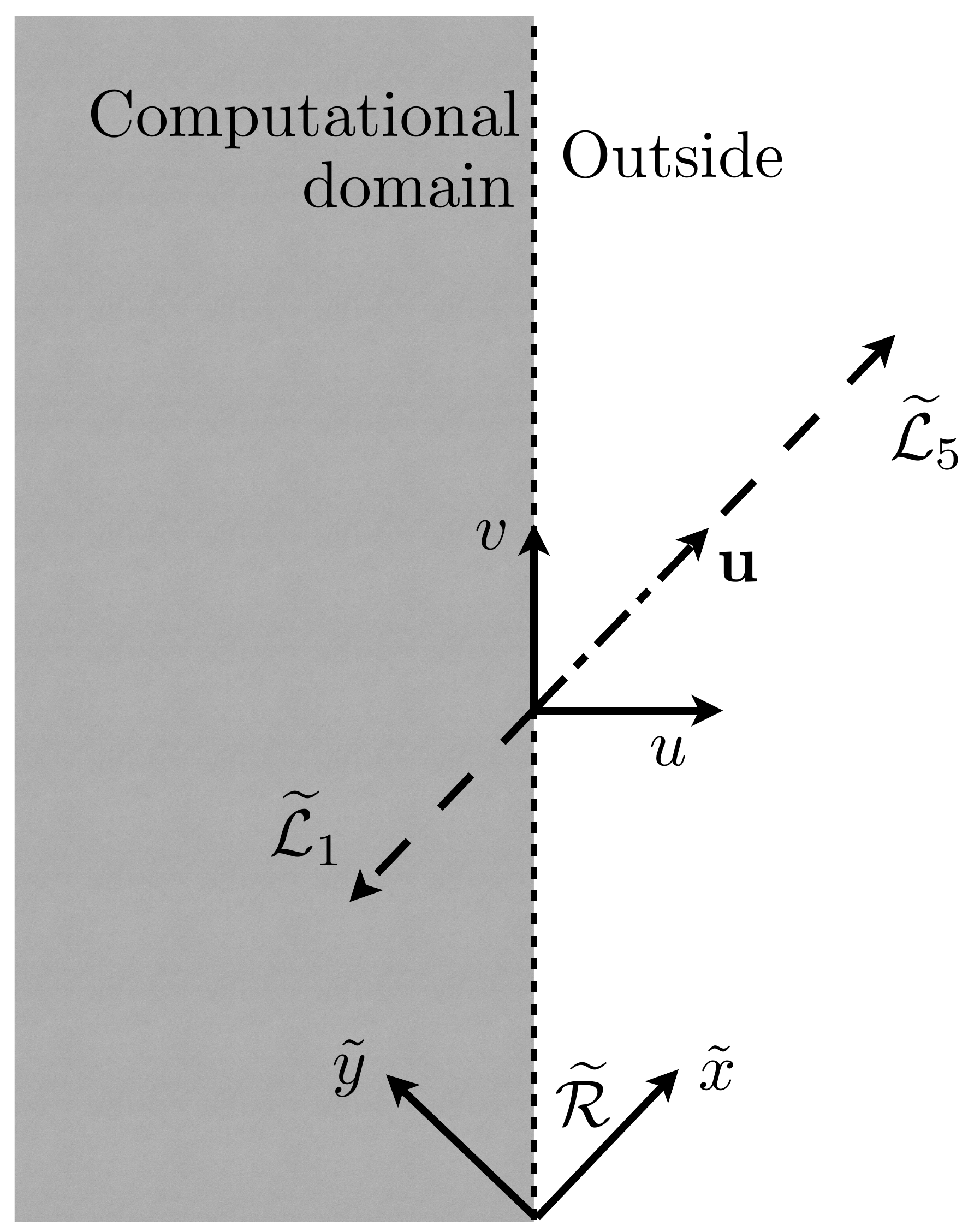}} \\
\caption{Schematic of characteristic wave projections in two coordinate systems : a) geometry based frame $R$. b) local streamline based frame $\tilde R$.\label{fig:Rtilde}}
\end{figure}

Another potential solution is to compute the LODI equation in the local streamline based frame $\tilde \mathcal{R}$ (Fig.~\ref{fig:Rtilde})~\cite{Albin_CompFluids_51_2011}. In order to compute the new characteristic waves $\tilde \mathcal{L}_i$, the non-dimensional velocity vector is projected into the new reference frame with the difficulty to compute $\tilde x$-derivative terms from the lattice discretization. A simple approximation is to set 
\begin{equation}
\frac{\partial \tilde \phi}{\partial \tilde x} = \frac{\partial \tilde \phi}{\partial x}, 
\end{equation}
and thus to compute it by a first-order upwind scheme using the lattice discretization. This implementation of the CBC condition will be referred to as the local streamline LODI (LS-LODI).

\subsection*{Adaptation to the Lattice Boltzmann scheme}

The main difficulty in LBM is then to find a set of populations in order to impose the physical values obtained by the CBC theory and the correct associated gradients. The possible adaptations can be divided in two families: those preserving the known particle populations (e.g. Zou/He BC~\cite{Zou_PhysFluids_9_1997}) and those replacing all particle populations (e.g. Regularized BC~\cite{Malaspinas_CompFluids_49_2011}). Izquierdo and Fueyo~\cite{Izquierdo_PhysRevE_78_2008} and Jung \textit{et al.}~\cite{Jung:2015} chose to modify the missing populations by adapting a pressure anti-bounceback boundary condition, which can be used as far as the MRT collision scheme is adopted~\cite{Ginzburg:2008}. Heubes \textit{et al.}~\cite{Heubes_JCAM_262_2014} decided to impose the CBC physical values thanks to the equilibrium populations, which is known to impose incorrect gradients at the boundary~\cite{Latt_PhysRevE_77_2008}. Three possibilities are introduced below and will be further evaluated: a modified Zou/He method and two declinations of the more stable regularized adaptation.

\begin{figure}[h!]
\begin{center}
\includegraphics[width=0.3\textwidth]{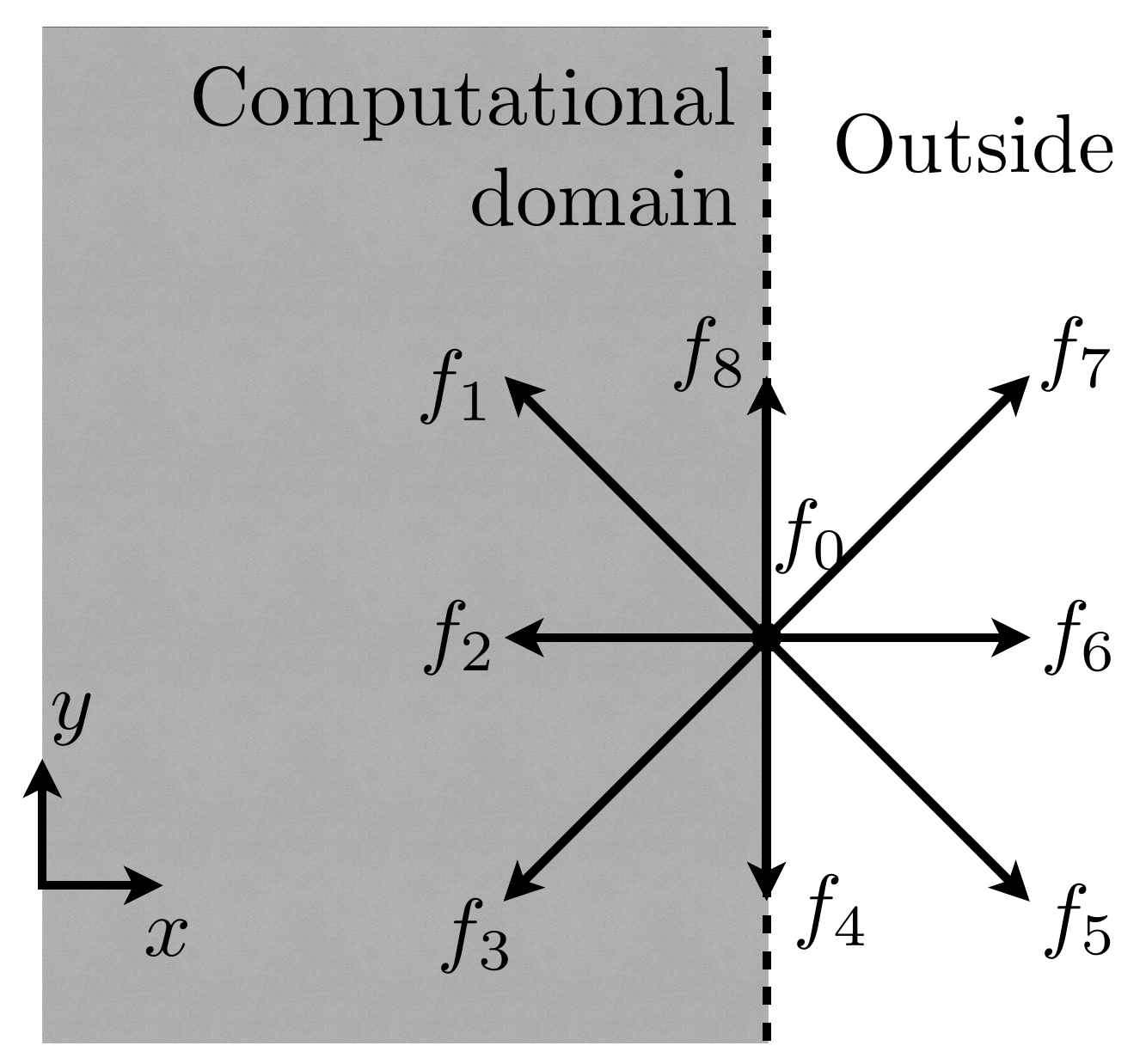}
\caption{Schematic plot of the population set at the boundary of a D2Q9 model. After the stream phase, $f_1$, $f_2$ and $f_3$ are missing.\label{fig:ZouHe}}
\end{center}
\end{figure}

\subsubsection*{Adaptation with Zou/He boundary conditions}
Let us consider an outflow boundary located at $x=L$  on a D2Q9 lattice (Fig.~\ref{fig:ZouHe}). After streaming, three incoming populations are missing : $f_1$, $f_2$ and $f_3$. 
The zeroth and first-order hydrodynamic moments are:
\begin{equation}
\rho = f_1+f_2+f_3+\rho_0+\rho_+,
\end{equation}
\begin{equation}
\rho u = \rho_+ - (f_1+f_2+f_3),
\end{equation}
where $\rho_+ = f_5+f_6+f_7$ and $\rho_0=f_0+f_4+f_8$. These equations can be combined, by eliminating the missing populations ($f_1+f_2+f_3$), to obtain
\begin{equation}
\label{RelationRhoU}
\rho = \frac{1}{1-u}(\rho_0+2\rho_+),
\end{equation}
where $\rho$ and $u$ are still non-dimensional. Thus, $\rho$ and $u$ are linked with relation (\ref{RelationRhoU}), which proves that it is impossible to impose both $\rho_b$ and $\mathbf{u_b}$ at the boundary without modifying any of the known populations.The same relation can be obtained for every 1D, 2D or 3D lattice as far as there is only one level of velocity. This relation concerns every boundary condition of the first family (those preserving the knwon populations) and will be used later. \newline

Let us note $g_i$ the corrected populations at the boundary. As proposed in~\cite{Zou_PhysFluids_9_1997}, the missing populations can be computed as follows:
\begin{equation}
\label{g1}
g_1=f_1^{(eq)} + f_5^{(neq)} + \frac{1}{2}(f_4^{(neq)} - f_8^{(neq)}),
\end{equation}
\begin{equation}
\label{g2}
g_2=f_2^{(eq)} + f_6^{(neq)},
\end{equation}
\begin{equation}
\label{g3}
g_3=f_3^{(eq)} + f_7^{(neq)} - \frac{1}{2}(f_4^{(neq)} - f_8^{(neq)}),
\end{equation}

while every other populations are kept unchanged: 
\begin{equation}
\label{fi}
g_i = f_i,\ \ \ \ \ \ \ \ i=0, 4, 5, 6, 7, 8,
\end{equation}

with $f_i^{(neq)}=f_i - f_i^{(eq)}$ and where the equilibrium populations are computed with the physical values imposed at the boundary, $\rho_b$ and $\mathbf{u_b}=(u_b, v_b)$. 
As shown in~\cite{Zou_PhysFluids_9_1997}, corrections (\ref{g1}), (\ref{g2}) and (\ref{g3}) then allow to impose the first-order moment at the boundary $\rho_b \mathbf{u_b}$. However, density and velocity are still linked with (\ref{RelationRhoU}). This is not a problem for a dirichlet Zou/He boundary condition where only one physical value (either $\rho$ or $\mathbf{u}$) is imposed and the other one is computed with (\ref{RelationRhoU}). On the contrary, for a non reflective outflow, both of them need to be imposed as the result given by the CBC method, so that the only condition set by the user is the value of a characteristic wave. It is then necessary to modify at least one known population. Schlaffer suggests correcting every populations in order to impose the correct density~\cite{Schlaffer:2013}. The solution proposed here is to add a correction on the population associated to a null velocity only:
\begin{equation}
\label{g0}
g_0=f_0 + \rho_b - \frac{1}{1-u}(\rho_0 + 2\rho_+),
\end{equation}
which ensures the value of $\rho_b$. This choice is motivated by the fact that this added correction will only affect the collision phase and will not be streamed into the computational domain. \newline

To sum up this method, $g_1$, $g_2$, $g_3$ and $g_0$ are computed with respectively (\ref{g1}), (\ref{g2}), (\ref{g3}) and (\ref{g0}) while other populations are kept unchanged after streaming:
\begin{equation}
\label{gi}
g_i=f_i,\ \ \ \ \ \ \ \ i=4, 5, 6, 7, 8.
\end{equation}

This method can be easily transposed on every 3D lattice (except for high order lattices~\cite{Philippi2006}) by using the general formula of the Zou/He boundary conditions that can be found in~\cite{Latt_PhysRevE_77_2008} for the missing populations, and correction (\ref{g0}) to ensure the value of $\rho_b$ and consequently $\mathbf{u_b}$.

\subsubsection*{Adaptation with the regularized method}
\label{regularized}
The Zou/He boundary condition provides the advantage of a very good precision in the definition of the boundary physical values. Unfortunately, this solution suffers from lack of stability for large Reynolds numbers. For example, it has been shown that Zou/He boundary conditions become unstable at $Re>100$ for a given resolution $N=200$ nodes per characteristic length in a 2D channel flow~\cite{Latt_PhysRevE_77_2008}. 
Another possible adaptation is to use the regularized boundary condition in order to impose the physical values computed by the CBC theory. This solution is yet less accurate but well more stable, as shown by Latt \textit{et al.} \cite{Latt_PhysRevE_77_2008}.\newline

More details about the regularized method for boundary conditions can be found for example in \cite{Latt_PhysRevE_77_2008}. The purpose
of this section is to explain how this particular boundary condition 
is used to impose $\rho_b$ and $\mathbf{u}_b$ on a flat boundary. 

The leading order of the populations $f_i$ can be expressed (see end of~(\ref{eq_fi_ce}) and~(\ref{eq_f1_pi1})) as
\begin{equation}
 f_i=f_i^{(eq)}(\rho_b,\mathbf{u_b})+f_i^{(1)}(\bm{\Pi^{(1)}}).\label{eq_f1_bc}
\end{equation}
On a boundary node the density $\rho_b$ and velocity $\mathbf{u}_b$ are imposed. Therefore in order 
to be able to use this last equation one needs a way to compute $\bm{\Pi^{(1)}}$. This is achieved
by using the fact that $\mathbf{Q}_i$ is a symmetric tensor with respect to $i$ which means that
$\mathbf{Q}_i=\mathbf{Q}_{opp(i)}$, where 
\begin{equation}
opp(i)=\{j|\mathbf{c}_i=-\mathbf{c}_j\}.
\end{equation}
At the leading order, this property leads to 
\begin{equation}
 f^{(1)}_i=f^{(1)}_{opp(i)}.
\end{equation}
The known $f_i^{(1)}$ can be straightforwardly computed by the following formula
\begin{equation}
f^{(1)}_i=f_i-f_i^{(eq)}(\rho_b,\mathbf{u}_b).
\end{equation}
With the last two equations, the set of $f_i^{(1)}$ is complete (they are all known) and can be used to compute $\bm{\Pi^{(1)}}$ through~(\ref{eq_pi1_f1}).
Then using~(\ref{eq_f1_pi1}) one recomputes regularized $f_i^{(1)}$ populations and the total populations $f_i$ are computed with the relation~(\ref{eq_f1_bc}). This method is valid in both 2D and 3D and will be called Regularized Bounceback (or Regularized BB) adaptation in the next sections. \newline
Another possibility is to compute $\bm{\Pi^{(1)}}$ by a second order finite difference scheme thanks to (\ref{pi1_grad}) and recompute $f^{(1)}$ using (\ref{eq_f1_pi1}). This method will be called the Regularized FD adaptation.

\subsection*{Summary of the method}

The non reflecting outflow boundary condition using CBC theory for LBM can be summarized as follows, considering everything is known at (non-dimensional) time $t$:
\begin{enumerate}
\item Computation of the physical values that must be imposed at $t+1$ to avoid non physical reflections by the CBC theory using either BL-LODI, CBC-2D or LS-LODI method. These values are stored to be used in the last step.\newline
\item Collision step.\newline
\item Streaming step: some populations are missing at the boundary.\newline
\item Correction of the set of populations at the boundary so that the physical values stored in the first step are imposed by using the Zou/He adaptation the so-called regularized Bounceback adaptation or the Regularized FD adapatation. \newline
\end{enumerate}

\section{Application to academic cases}

In this section, the CBC approach for LBM is assessed on simple 2D cases: a normal plane wave, a plane wave with different incidence angles, a spherical density wave and a convected vortex.

\subsection*{2D normal plane wave}
\begin{figure}[h!]
\includegraphics[width=\linewidth]{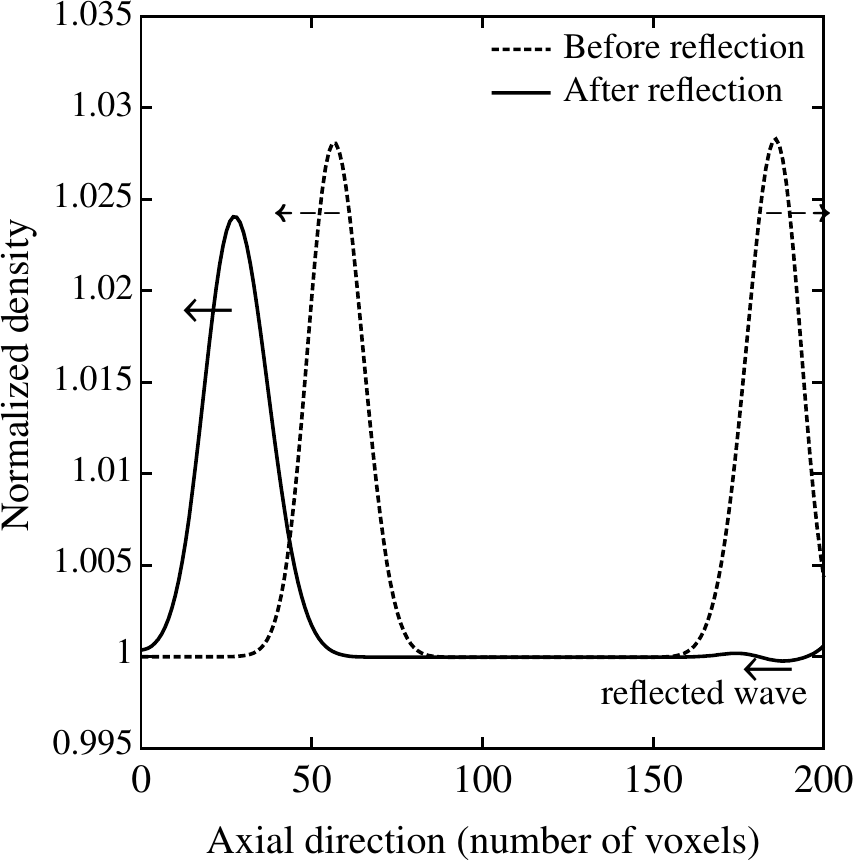}
\caption{Density waves before and after reflection of the $(u+c_s)$ wave on the right boundary (non-reflecting outflow).\label{fig:rho4_8bis}
}
\end{figure}

The computational domain, a square of $200\times200$ cells, is initiated with a gaussian plane wave as follows (in lattice units)
$$\rho_0=1+0.1*\exp\left(-\frac{(x-x_0)^2}{2\sigma^2}\right),$$
$$u_0=0.1,$$
$$v_0=0.1*\exp\left(-\frac{(x-x_0)^2}{2R_c^2}\right),$$
with $x_0=110$ and $2R_c^2=20$ in lattice units (\textit{i.e.} in number of cells). The Reynolds number, computed with the horizontal non-dimensional initial velocity $u_0$, the characteristic size of the box in number of voxels and the viscosity in lattice units, is equal to $100$.
The boundary conditions are: vertical periodicity, reflecting inlet on the left and perfectly non reflecting outflow on the right. For this pure 1D case, the choice of the CBC condition (baseline, local streamline LODI or LODI with transverse terms) has no effect on the absorption rate. Moreover, all the three adaptations provide the same results at $10^{-6}$ and no stability issues have been encountered with the Zou/He adaptation for this test case. Thus, the results presented below have been obtained with baseline LODI and Zou/He adaptation.
This test case has been chosen so that the reflection rate for every macroscopic value ($\rho$, $u$ and $v$) can be computed, since two pressure and axial velocity waves propagates at speed $(u-c_s)$ and $(u+c_s)$ and one transverse velocity wave propagates at speed $u$. The computed density waves are represented on Fig.~\ref{fig:rho4_8bis} at two different time steps: before reflection of the $(u+c_s)$ wave on the non-reflective outlet and shortly after reflection. 

A very low reflected amplitude can be distinguished. The $(u-c_s)$ wave propagates to the left of the domain without encountering any boundary at the two observed time steps. It is only affected by viscous dissipation and is used as a reference amplitude in the computation of the reflection rates of density and axial velocity. For the transverse velocity wave, one can compute the reflection rate as the ratio between the reflected wave amplitude and the amplitude shortly before reflection. The obtained results are presented on Table~\ref{table:ReflectionRate_PlaneWave}.

\begin{table}[h!]
\begin{center}
\begin{tabular}{|c| c| c| c|}
\hline
   Variable & $\rho$ & $u$ & $v$ \\
\hline
   Reflection rate & 1.2 \% & 1.1 \% & $<10^{-5}$ \% \\
\hline
\end{tabular}
\caption{Reflection rates for the 2D normal plane wave. \label{table:ReflectionRate_PlaneWave}}
\end{center}
\end{table}

As often with CBC, the treatment is more difficult for the pressure wave, but the results obtained here are in good agreement compared to what is found in the literature \cite{Izquierdo_PhysRevE_78_2008, Heubes_JCAM_262_2014}. 

It is noticed that $\rho=1$ is not correctly recovered at the outlet after reflection, as the boundary has been set as perfectly non-reflecting ($\mathcal{L}_1=0$). In order to impose the correct boundary condition, a relaxation should be implemented, as in eq (\ref{L1}). However, in that case, the reflection rate will increase as shown in~\cite{Poinsot_JCP_101_1992}.

\subsection*{2D plane wave with incidence}

At $t=0$, the computational domain is at rest ($\rho_0=1$, $\mathbf{u}_0=\mathbf{0}$) except for an oblique line on which the density is set to $\rho_{0}=1.1$ in order to generate a plane wave with an incidence $\alpha$. The reflection coefficient is measured by computing the ratio of maximal amplitudes in density waves only, as this is the most critical hydrodynamic variable. The three implementations are tested for this case: BL-LODI, LS-LODI and CBC-2D in the perfectly non-reflecting case. Again, the results obtained with the Zou/He adaptation only will be presented here, as the same results at $10^{-6}$ have been obtained with the Regularized BB and Regularized DF adaptations.

\begin{SCfigure}[][h!]
\begin{overpic}[width=0.35\textwidth]{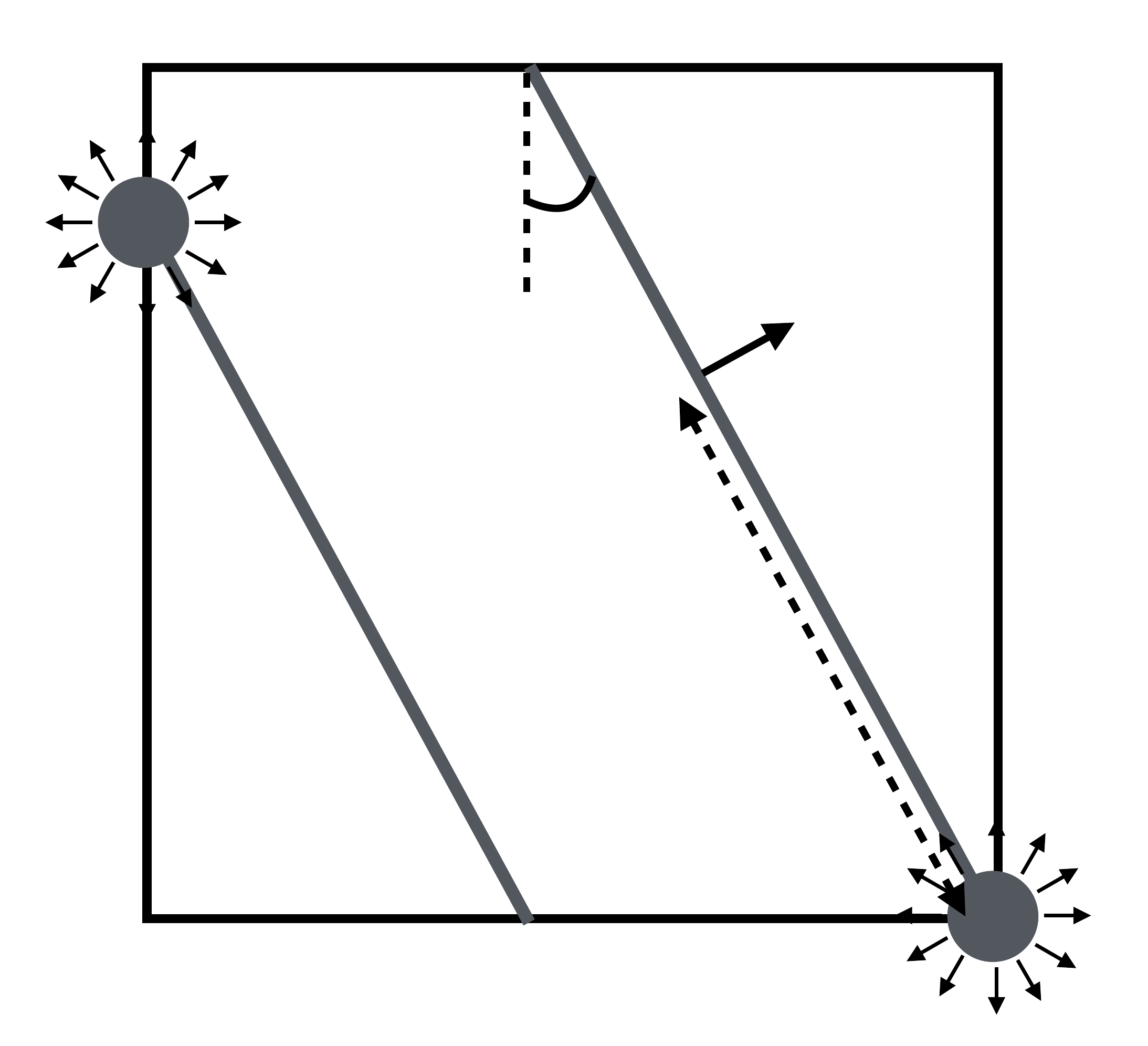}
\put(49,68){$\alpha$}
\put(53,56){$M$}
\put(63,65){$c_s$}
\put(65,35){$h$}
\put(90,75){\rotatebox{270}{ \small Non-reflective outlet}}
\put(5,20){\rotatebox{90}{ \small Reflective inlet}}
\put(40,88){\small Periodicity}
\put(40,5){\small Periodicity}
\end{overpic}
\caption{Schematic plot of the initialization of the plane wave test case with an incidence angle $\alpha$. Spherical waves instantaneously appear at the two extremities to distort the plane wave.} \label{fig:ObliqueWaveProblem}
\end{SCfigure}
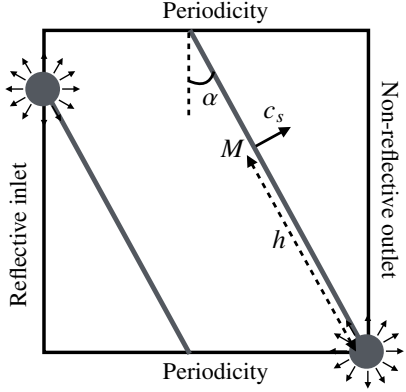

Because of a spurious phenomenon appearing at high incidence angles, only angles below $45 ^{\circ} $could be computed. Indeed, contrary to the previous normal wave test case, the incident wave cannot be infinite in its transverse direction. Then, spherical waves appear at each extremity of the oblique line initializing the wave, as shown on Fig.~\ref{fig:ObliqueWaveProblem}. Let us imagine a point $M$ located at a distance $h$ of one extremity and moving with the oblique plane wave at velocity $c_s$. The spherical wave reaches $M$ after a time $h/c_s$. The point $M$ reaches the non-reflective outlet boundary condition at a time $h \tan (\alpha)/{c_s}$. The condition for this point of the wave to reach the outlet before being distorted is: 
\begin{equation}
\frac{h}{c_s}\tan(\alpha) < \frac{h}{c_s} \Leftrightarrow \tan(\alpha) < 1,
\end{equation}
which means than only incidence angles below $45^\circ$ can be computed with such a test case. This problem is avoided in~\cite{Heubes_JCAM_262_2014} by imposing an 'exact' solution obtained on a larger computational domain until the desired wave reaches the boundary. The exact solution is then switched to the CBC condition. It has not been tested in this paper in order to avoid the eventual acoustic waves generated by a brutal change in the boundary condition.\newline

\begin{figure}[h!]
\includegraphics[width=\linewidth]{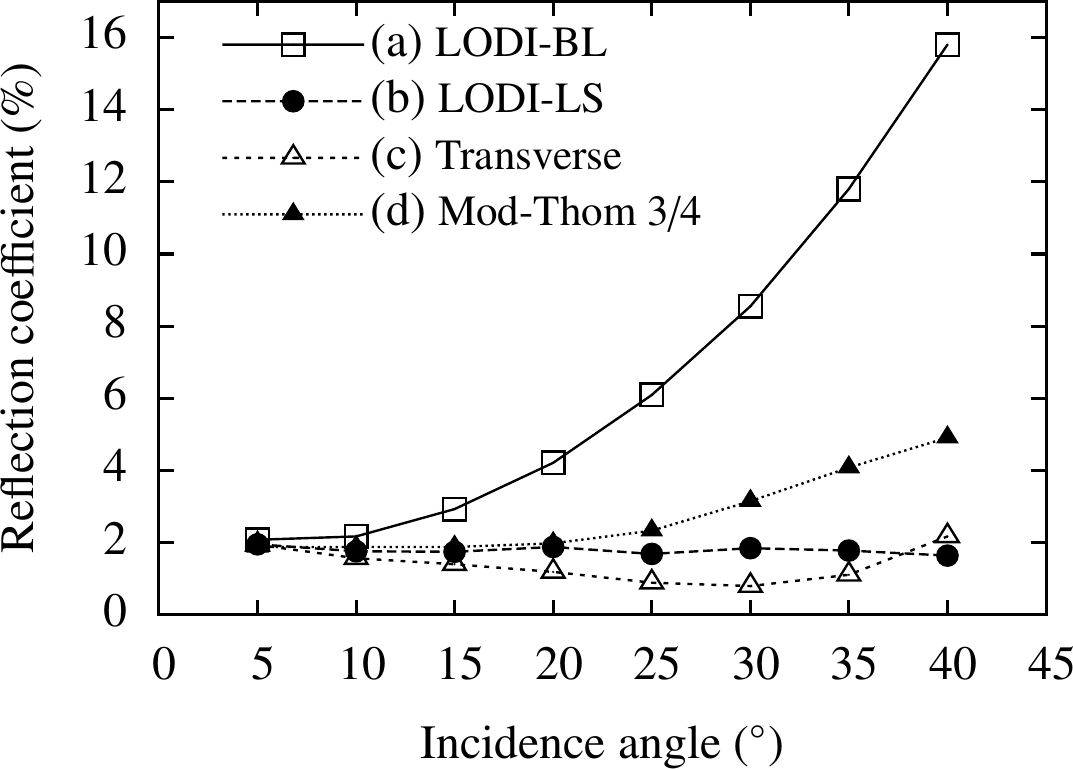}
\caption{Reflection coefficient (\%) with respect to the incidence of the plane wave: (a) baseline LODI, (b) local streamline LODI, (c) CBC-2D~\cite{Jung:2015} and (d) Mod-Thompson 3/4~\cite{Heubes_JCAM_262_2014}.\label{fig:Reflexion_incidence}}
\end{figure}

The reflection coefficient of the density wave with respect to the incidence angle is represented on Fig.~\ref{fig:Reflexion_incidence} for BL-LODI, LS-LODI and CBC-2D. The results are compared with what is obtained for the same test case with the modified Thompson method with a coefficient $\gamma=3/4$ as introduced in~\cite{Heubes_JCAM_262_2014}.

As expected, for the baseline approach (a), the reflection coefficient increases as the incidence angle increases. For the modified Thompson approach, the results are close to what can be found in~\cite{Heubes_JCAM_262_2014}: the reflection rate slightly increases and reaches $5\%$ at $40^\circ$ of incidence. On the contrary, when the CBC method is extended with the effects of transverse waves as in~\cite{Yoo:2007, Jung:2015}, the reflection rate decreases until $30^\circ$ and then begins to increase. In the case of a local streamline LODI implementation, the coefficient remains stable at around $2\%$ of reflected wave.

\subsection*{2D Spherical wave}

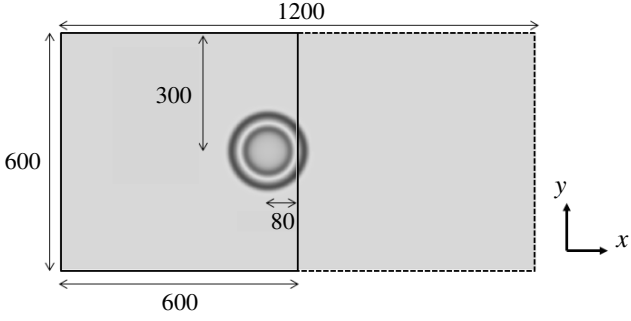
\begin{figure}[h!]
\begin{overpic}[scale=0.45]{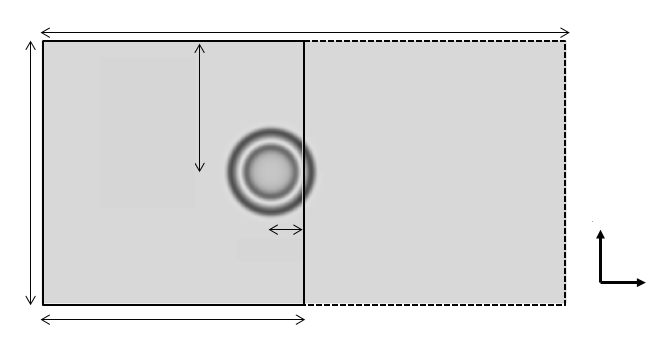}
\put(23,-0.5){\small 600}
\put(-3,23){\small 600}
\put(42,48){\small 1200}
\put(22,34){\small 300}
\put(41,13){\small 80}
\put(88,19){$y$}
\put(98,10){$x$}
\end{overpic}
\caption{Schematic plot of the computational domain - solid line: CBC cases (a), (b) and (c), dotted line: reference case (d). Dimensions are given in number of cells.\label{fig:Testcase_SphericalWave}}
\end{figure}

The computational domain, a square of $600\times600$ cells is initiated with a gaussian density in order to generate a spherical wave, as follows :
$$\rho=1+0.1*\exp\left(-\frac{((x-x_0)^2+(y-y_0)^2)}{2R_c^2}\right),$$ 
with $R_c = 3.2$, $x_0=520$ and $y_0=300$ nodes.
The boundary conditions are periodic in the vertical direction, a CBC condition is set on the right boundary and a reflective boundary condition on the left (located far enough to avoid its influence on the spherical wave at the studied time steps). Four cases are computed: (a) baseline LODI, (b) CBC with transverse terms, (c) local streamline LODI and (d) reference case where the domain is enlarged in the horizontal direction so that the spherical wave is not affected by the boundary (Fig.~\ref{fig:Testcase_SphericalWave}). As for the previous test cases, the LBM adaptation of the CBC condition had no impact on the results.

The local reflection coefficient for such a spherical wave can be computed by the following formula:
\begin{equation}
r = \frac{R-A_{ref}}{|I|},
\end{equation}
where $I$ is the amplitude of the original density wave running towards the boundary condition, $R$ is the amplitude of the density wave after reflection at the outlet and $A_{ref}$ is the amplitude of the density wave at the same lattice node compared to the reference case (d) (no reflection). 

\begin{figure}[h!]
(a) \\
 \includegraphics[scale=0.45]{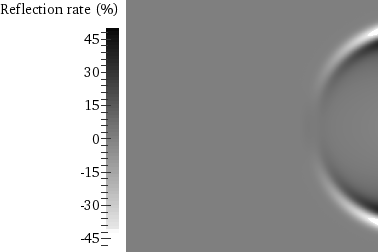}\\
(b)\\
\includegraphics[scale=0.45]{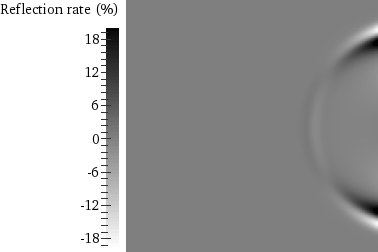}\\
(c)\\
\includegraphics[scale=0.45]{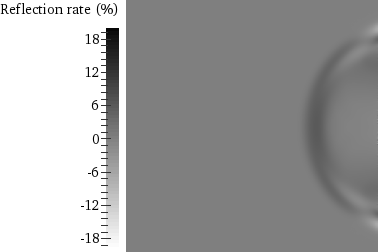}\\
\caption{Density wave after reflection of a spherical wave at the outlet (400 iterations): (a) baseline LODI, (b) CBC-2D~\cite{Yoo:2007, Jung:2015} and (c) local streamline CBC.\label{fig:density_22_R}}
\end{figure}

A map of reflection rates after reflection at the outlet (after 400 iterations) is shown on Fig.~\ref{fig:density_22_R}. For the baseline approach, one can see that the reflected wave becomes more important as the local incidence of the spherical waves increases, because the incident wave can be approximated as a local normal plane in the center of the outlet while, in the corners, the incidence becomes important (up to $75^{\circ}$). When adding transverse term without any relaxation as in~\cite{Yoo:2007, Jung:2015}, the observation confirms the prospective behavior of the plane waves simulations: even if the reflection rate is consistently reduced for angles below $40^\circ$, it reaches $20\%$ for the greatest observed angles. On the contrary, the reflection rate remains constant with the local streamline LODI implementation.




\subsection*{2D convected vortex}
A 2D vortex is convected from left to right and exits the computational domain, a square of $600\times600$ grid points. A particular attention must be paid to the initialization of the Lamb-Oseen vortex~\cite{Lamb:1932} which has to be adapted to the isothermal case in order to avoid spurious waves generated by the adaptation of a wrong initial density. The initial conditions, in lattice units, are imposed as follows:

\begin{equation}
u = u_0 - \beta u_0 \frac{(y-y_0)}{R_c} \exp \left( -\frac{r^2}{2R_c} \right),
\end{equation}
\begin{equation}
v = \beta u_0 \frac{(x-x_0)}{R_c} \exp \left( -\frac{r^2}{2R_c} \right), 
\end{equation}
\begin{equation}
\rho = \left[ 1- \frac{(\beta u_0)^2}{2C_v} \exp \left( -\frac{r^2}{2} \right) \right] ^{1/(\gamma-1)},
\label{RhoVortex}
\end{equation}

where $u_0=0.1$ in lattice units, $\beta=0.5$, $x_0=y_0=300$ nodes (the vortex is initially centered on the box), $R_c=20$ nodes and $r^2=(x-x_0)^2+(y-y_0)^2$. With the BGK collision operator with a single relaxation time, the simulated gas has the following constants~\cite{Guo:2007}:
\begin{equation}
\gamma=\frac{D+2}{D},
\end{equation}
\begin{equation}
C_v=\frac{D}{2} c_s^2, 
\end{equation}
where $D=2$ is the dimension of the problem. The specific heat capacity at constant volume $C_v$ appears instead of $C_p$ in (\ref{RhoVortex}) because of an error in the heat flux obtained in the Navier-Stokes equations after the Chapman-Enskog development for athermal lattices~\cite{Guo:2007}. The Reynolds number based on $u_0$ and the size of the computational domain is equal to 1000 and the Regularized BGK scheme is chosen for the collision step. This convected vortex test case is a well known test for boundary conditions as it often reveals spurious distorsions at boundaries due to the presence of transverse terms in the Navier-Stokes equation~\cite{Yoo_CTM_11_2007}. As previously, top and bottom boundary conditions are periodic, the left condition is a regularized inlet and four different CBC conditions will be evaluated at the right boundary:\newline
\indent (a) baseline LODI with $K_1=0$,\\
\indent (b) local streamline LODI with $K_1=0$,\\
\indent (c) CBC-2D with $\mathcal{L}_1=\mathcal{T}_1$,\\
\indent (d) CBC-2D relaxed including transverse terms with: $$\mathcal{L}_1=\frac{\sigma (1-M^2)c_s^3}{R_c}(\rho-1) + (1-M)\mathcal{T}_1,$$ where $M=0.2$ and $\sigma=0.9$.\\
For low Reynolds numbers ($Re < 1000$), simulations --- not shown here --- provided the same results at $10^{-6}$. However, for $Re=1000$, the Zou/He adaptation was no more stable for this test case, contrary to both regularized methods. The results presented here have been obtained with the Regularized BB adaptation.

\begin{figure}[h!]
\begin{center}
(a) Baseline LODI\ \ \ \ \ \ \ \ \  (b) LS LODI\ \ \ \ \ $ $
\includegraphics[width=0.2\textwidth]{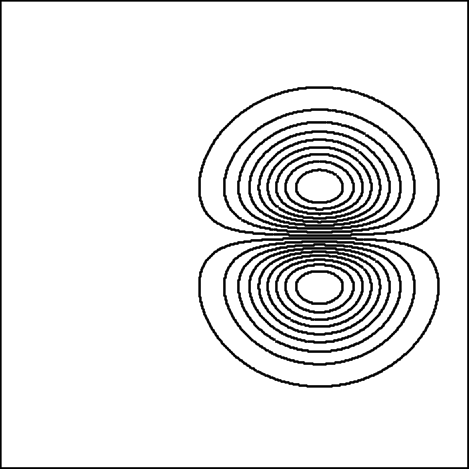} \includegraphics[width=0.2\textwidth]{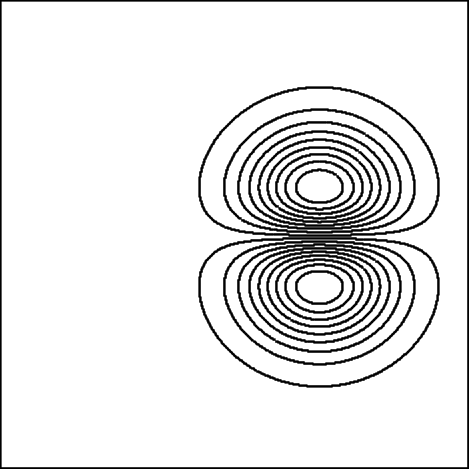} \\

\includegraphics[width=0.2\textwidth]{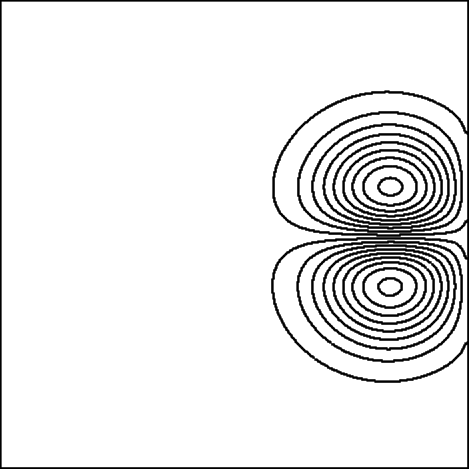} \includegraphics[width=0.2\textwidth]{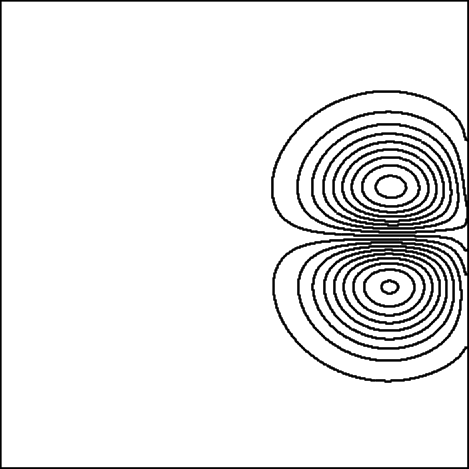} \\

\includegraphics[width=0.2\textwidth]{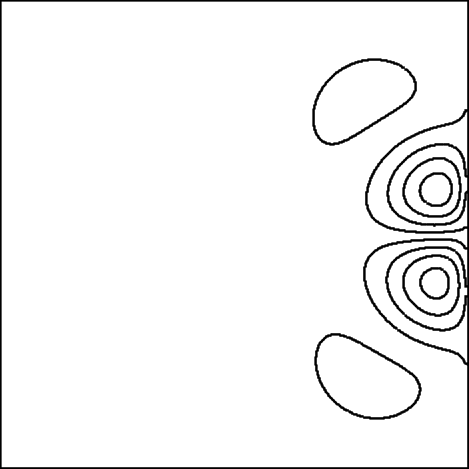} \includegraphics[width=0.2\textwidth]{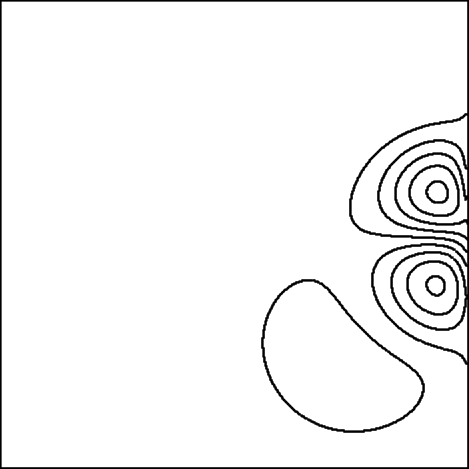} \\

\includegraphics[width=0.2\textwidth]{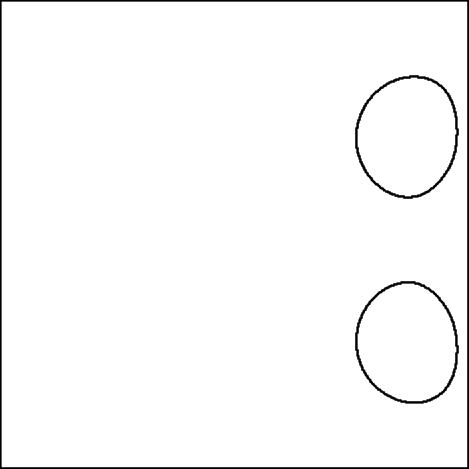} \includegraphics[width=0.2\textwidth]{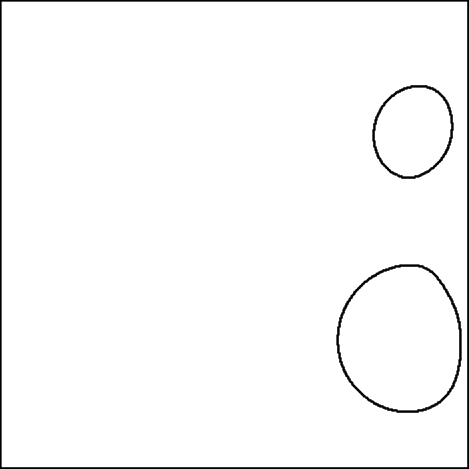} \\
\caption{Isovalues of longitudinal velocity $(u-u_0)/u_0$ (min=-0.285, max=0.285) of the convected vortex leaving the computational domain at four time steps: 400, 700, 1000 and 1300 iterations. (a) Baseline LODI and (b) LS-LODI. A small part of the computational domain is represented. \label{fig:CoVo_IsoV}}
\end{center}
\end{figure}

\begin{figure}[h!]
\begin{center}
\ \ \ \ \ \ \ \ (c) CBC-2D\ \ \ \ \ \ (d) CBC-2D relaxed\ \ \ 
\includegraphics[width=0.2\textwidth]{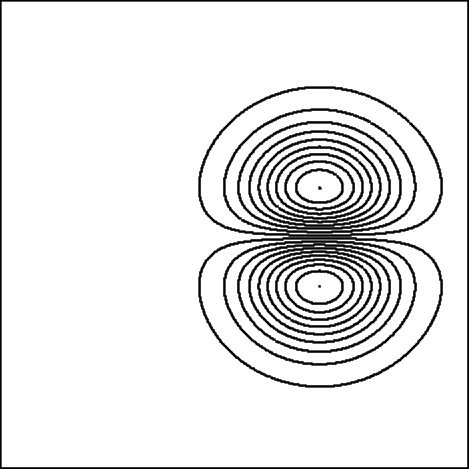} \includegraphics[width=0.2\textwidth]{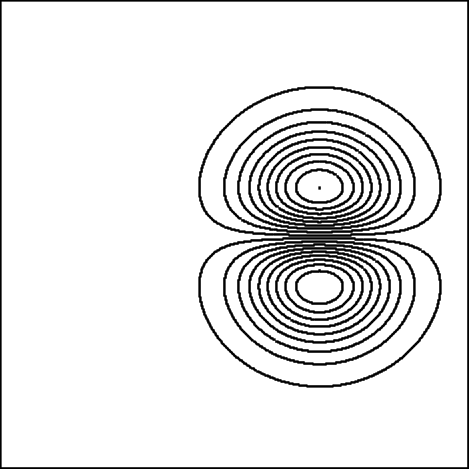} \\

\includegraphics[width=0.2\textwidth]{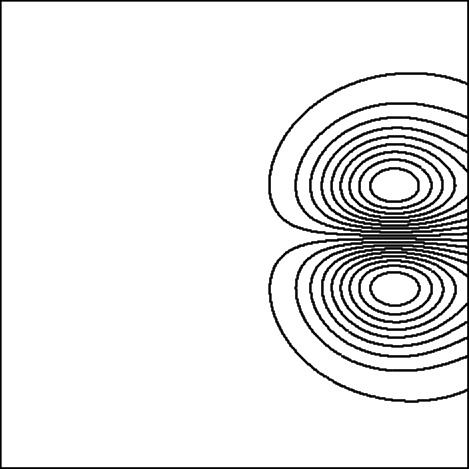} \includegraphics[width=0.2\textwidth]{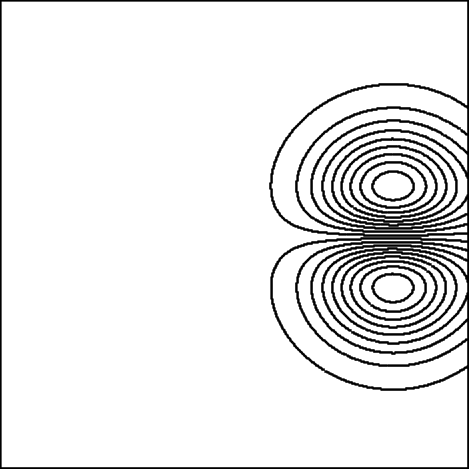} \\

\includegraphics[width=0.2\textwidth]{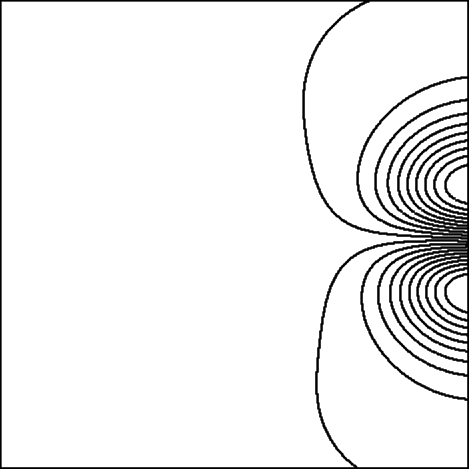} \includegraphics[width=0.2\textwidth]{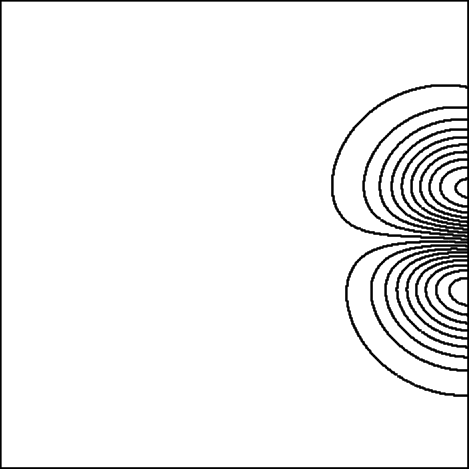} \\

\includegraphics[width=0.2\textwidth]{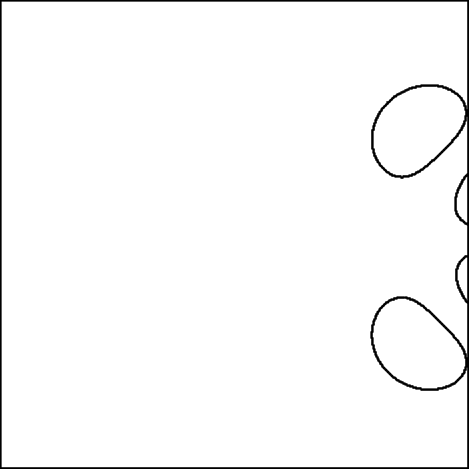} \includegraphics[width=0.2\textwidth]{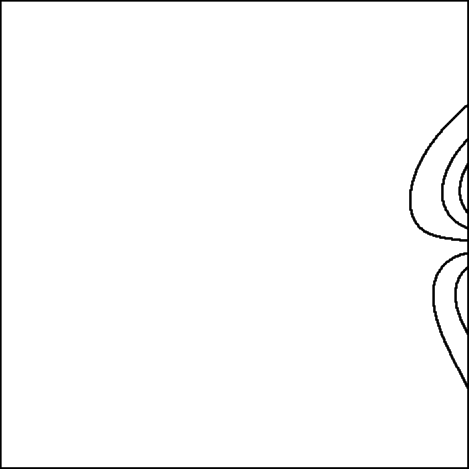} \\
\caption{Isovalues of longitudinal velocity $(u-u_0)/u_0$ (min=-0.285, max=0.285) of the convected vortex leaving the computational domain at four time steps: 400, 700, 1000 and 1300 iterations. (c) CBC-2D and (d) CBC-2D relaxed. A small part of the computational domain is represented. \label{fig:CoVoTransverse_IsoV}}
\end{center}
\end{figure}

Fig.\ref{fig:CoVo_IsoV} and \ref{fig:CoVoTransverse_IsoV} show isovalues of non dimensional longitudinal velocity for the four studied cases. First, it can be noticed that, contrary to what was observed in the previous test cases, the use of local streamline boundary condition does not allow to reduce non physical reflections compared to baseline LODI. A possible explanation would be that, inside the vortex, local streamlines are nearly perpendicular to the direction of propagation of the local wave, whereas they were aligned in the case of a pure acoustic wave. The LODI equations are thus applied in a wrong frame which generates non physical reflections. Results are slightly better for the BL-LODI for which the local frame is the good one for at least local normal waves. The addition of the transverse waves in the CBC-2D boundary allows the vortex to keep a correct shape at least until 1000 iterations. However, once the first half of the vortex has reached the outlet, it is distorted as in the BL-LODI case. The addition of relaxation coefficients allows the vortex to keep its shape at the last iteration, even if it is a bit distorted.
It can be noticed that the configuration appears to be symmetric for boundaries (a) and (c), which is not the case for (b) and (d). Indeed, with the LS-LODI, the streamlines used in the change of frame are not symmetric and in the last case, the density frame appears to be asymmetric which has an influence on $\mathcal{L}_1$ through the relaxed pressure, and thus on the longitudinal velocity fields. 

\subsection*{Stability analysis}

An analysis of numerical stability of the previous convected vortex test case at different Reynolds numbers and different grid resolutions has been carried out. It has been noticed that the CBC type (BL-LODI, LS-LODI or CBC-2D) had no impact on the stability as far as $\mathcal{L}_1$ is not relaxed: the numerical stability comes from the LBM adaptation itself.

\begin{figure}[h!]
\includegraphics[width=\linewidth]{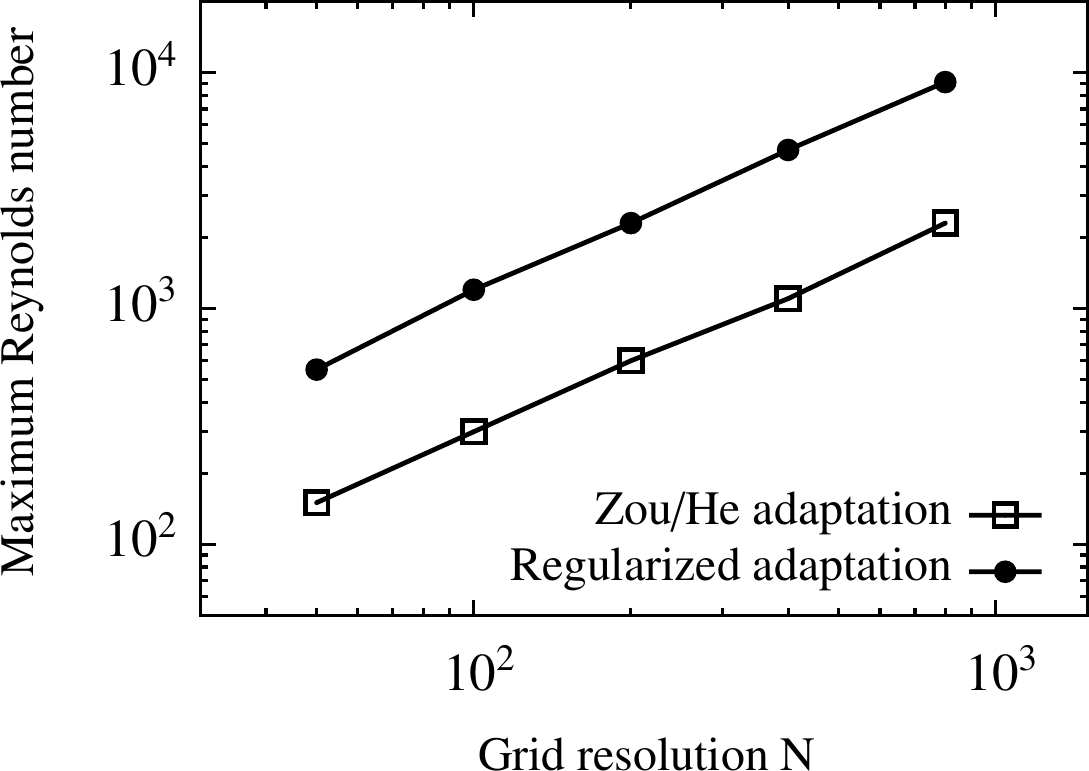}
\caption{Numerical stability of the Zou/He and Regularized BB adaptation of the CBC for the 2D convected vortex case. The maximum Reynolds number which can be reached before numerical instabilities appear is plotted. Regularized FD is not presented here since it has shown no numerical instability.\label{fig:stability}}
\end{figure}

Fig.~\ref{fig:stability} compares the stability of the Zou/He adaptation and the Regularized BB adaptation. As predicted from~\cite{Latt_PhysRevE_77_2008}, the classical regularized boundary condition is more stable than the Zou/He adaptation. However, a huge resolution is still required in order to reach high Reynolds numbers. On the contrary, the Regularized FD adaptation has shown to be unconditionnally numerically stable: in every configurations of the convected vortex test case, the first numerical instabilities came from the inside of the domain and not the CBC boundary condition.

\section{Application to a NACA0015 profil at high Reynolds number}
\begin{figure}
\begin{center}
\includegraphics[width=0.4\textwidth]{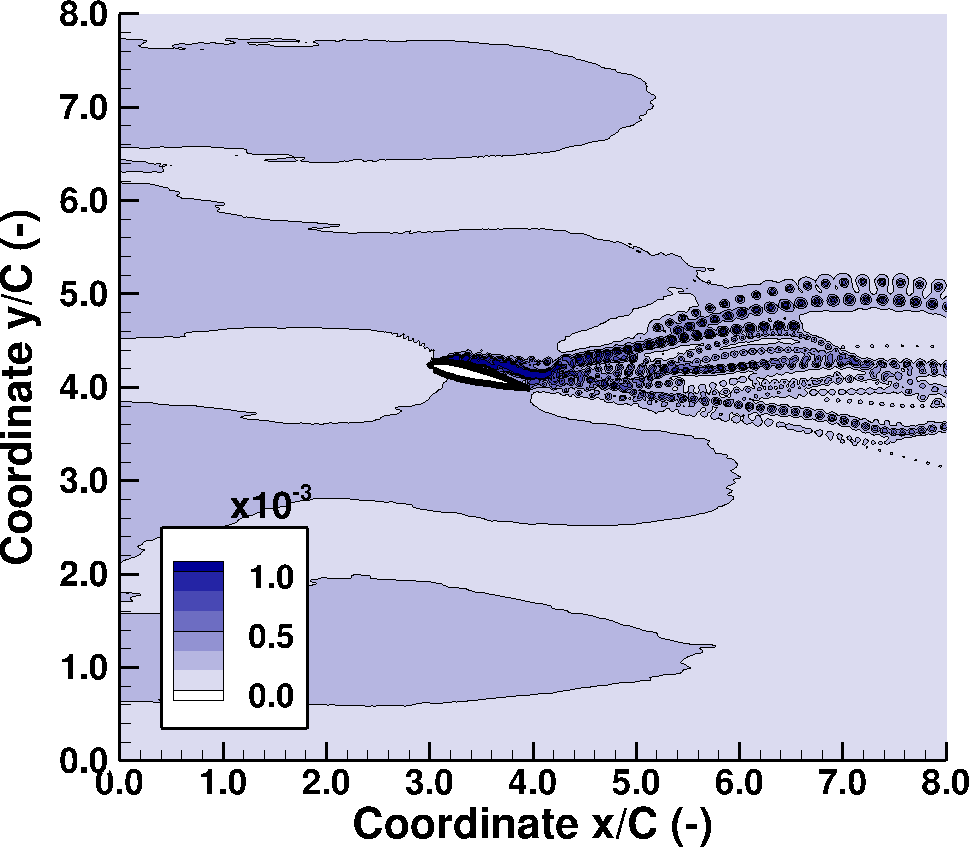}
\caption{Simulation of the flow around a NACA0015 profile at $Re=10^5$: time-averaged flow field colored with the pressure fluctuation $p_{RMS}$ (reference configuration, 16$C\times8C$ domain, only the left half of the domain is representedS). \label{fig:NACA1_prms_ref}}
\end{center}
\end{figure}

The objective of this section is to demonstrate the robustness of the CBCs in a case relevant for high Reynolds aerodynamics applications. The configuration is a NACA0015 profile, in a 8$C\times8C$ domain (with $C=1$~m the chord of the profile). The Mach number is set to 0.04 and the Reynolds number is set to $Re=10^5$. The lattice dimension is $\Delta x=1/400$~m, corresponding to a time step $\Delta t=4.25\times10^{-6}$~s. Each simulation is run for 200 000 time steps, but only the last 100 000 time steps are kept for data post-processing. The simulations are performed on a $3,200\times3,200$ points 2D grid, to ensure numerical stability and a proper resolution of the flow patterns with a LES formalism. The sub grid scale model is the Smagorinsky model, with a constant $C_s=0.18$, associated with a Regularized BGK collision scheme. In order to generate complex flow patterns, the profile is inclined with an angle of $15^o$, compared to the freestream velocity direction which is purely axial. Three CBCs are tested: BL-LODI, LS-LODI and CBC-2D. The CBC solutions are compared with a reference solution obtained on 16~$C\times8C$ domain, Fig.~\ref{fig:NACA1_prms_ref}. In this reference case, vortices do not leave the extended computational domain at the observed time steps and the initial acoustic wave is evacuated thanks to a Neumann outlet (populations at the boundary are copied from the neighbor voxel) associated with a 4C-wide viscosity sponge zone inside which the relaxation time is increased up to $\Delta t$ with a sinusoidal shape in order to smoothly increase the numerical dissipation. Among CBC computations, only the regularized FD adaptation has been sufficiently robust to achieve the simulations.

At this Reynolds number, the real flow should be 3D in the vicinity of the profile and in the wake, due to the development of turbulent flow patterns. However, the present 2D approach is not able to reproduce such 3D effects. Despite that, the trajectory of the vortices exhibits a chaotic behavior, as observed on the time-averaged flow field in Fig.~\ref{fig:NACA1_prms_ref}. The challenges with the present test case are to ensure that:
\begin{itemize}
\item{the initial acoustic wave generated by the presence of the airfoil leaves the domain with minimum reflection},
\item{the vortices generated by the boundary layer separation are correctly convected beyond the outlet plane with minimum spurious wave reflection}. 
\end{itemize}

\begin{figure}
\begin{center}
(a) \\
\includegraphics[width=0.4\textwidth]{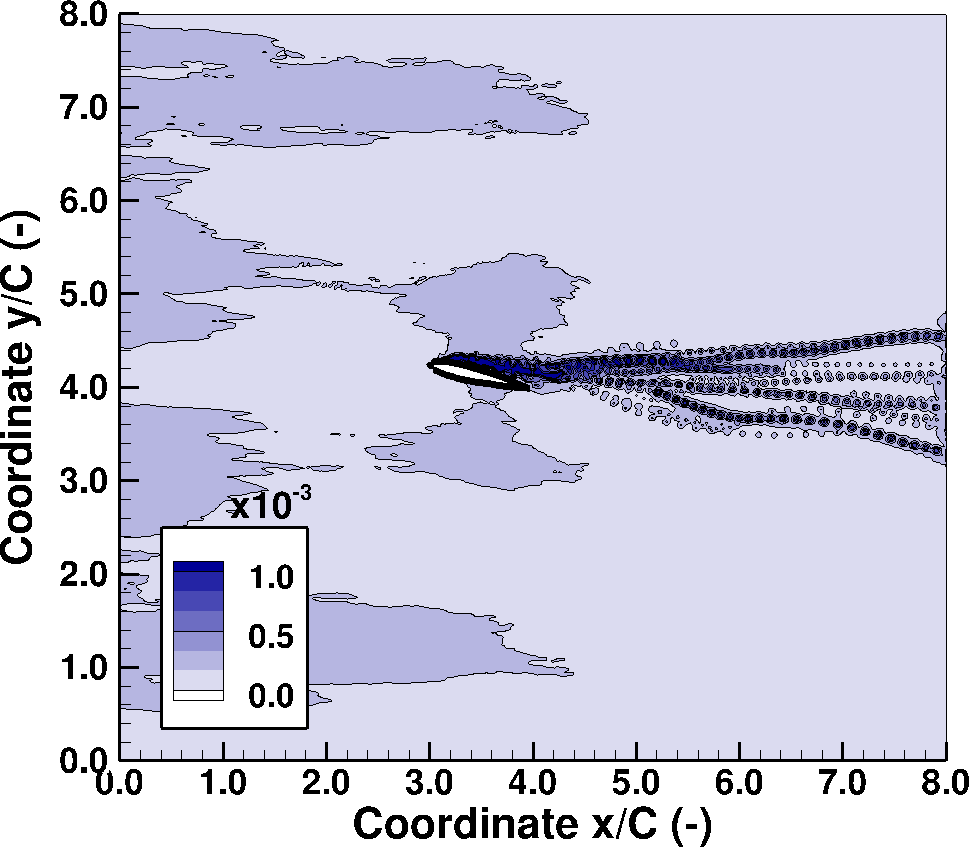}\\
(b) \\
\includegraphics[width=0.4\textwidth]{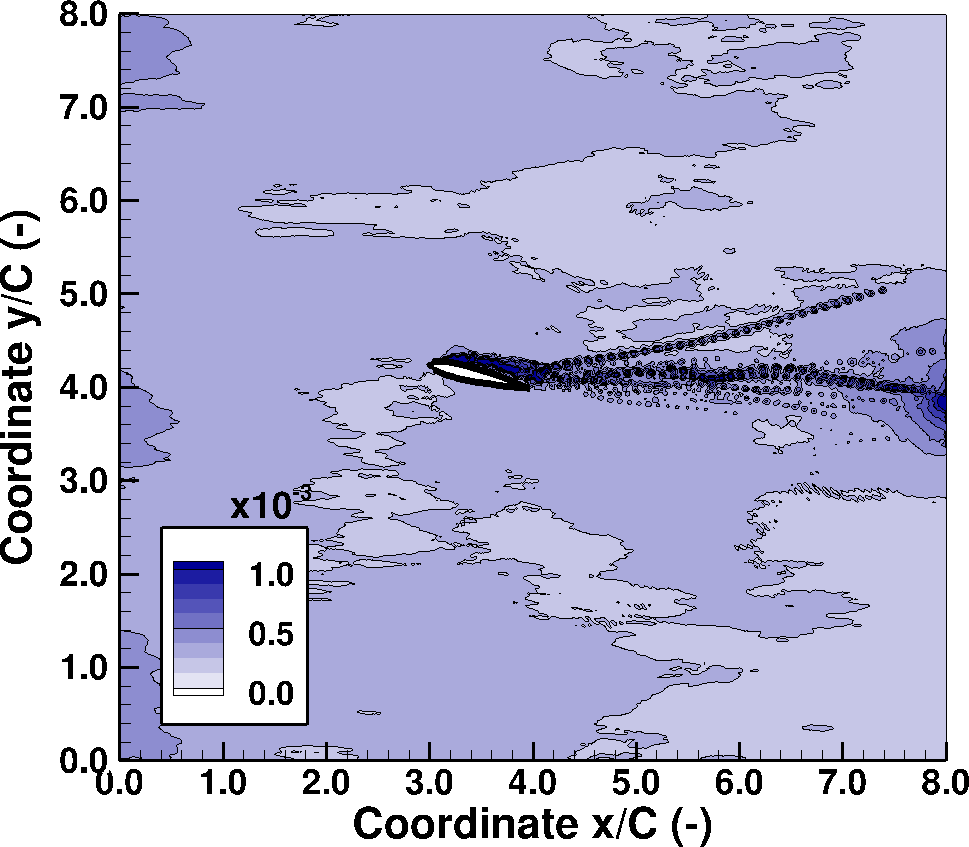}\\
(c) \\
\includegraphics[width=0.4\textwidth]{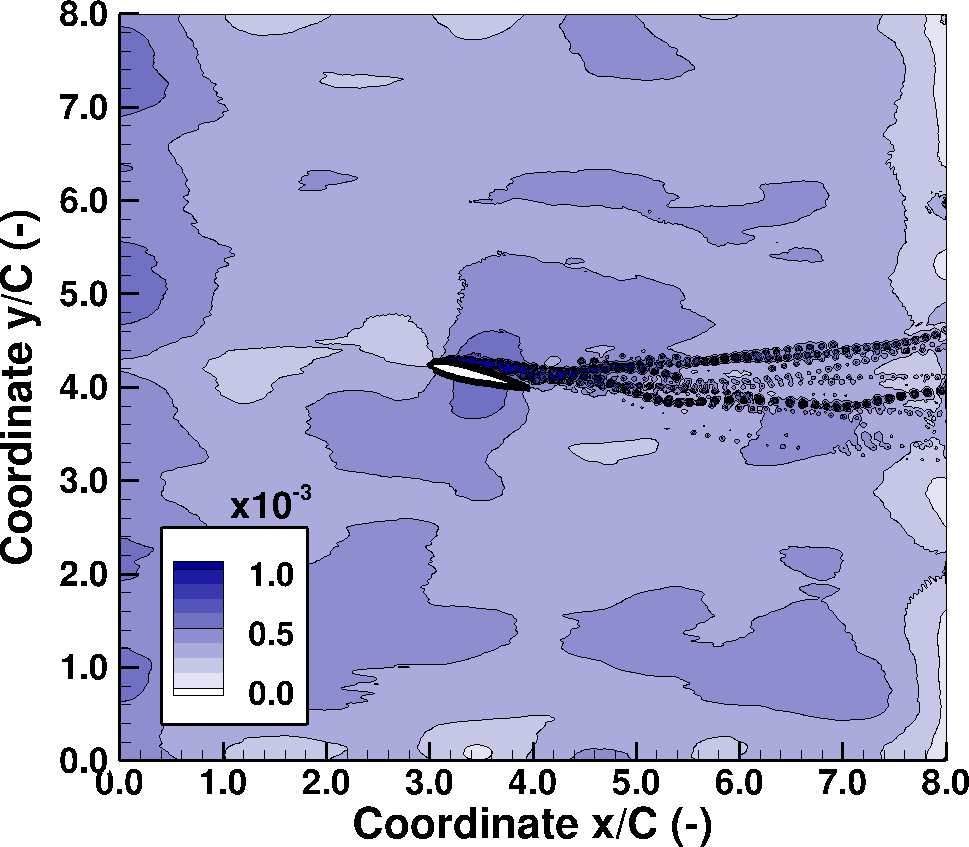}\\
\caption{Simulation of the flow around a NACA0015 profile at $Re=10^5$: time-averaged flow field colored with the pressure fluctuation $p_{RMS}$. (a) BL-LODI, (b) LS-LODI, (c) CBC-2D. \label{fig:NACA1_prms_CBC}}
\end{center}
\end{figure}

The difficulty for such a test case lies in the chaotic nature of the flow, since the trajectory of the vortices is affected by small perturbations. Because of this phenomenon, the visualization of error fields, computed as the difference between the reference solution and each tested CBC, would not be compellant since vortices are not superposed in each computation. A better choice is to plot the root mean square of the pressure field, defined as $p_{RMS}=\sqrt{\overline{p'^2}}/{\overline{p}}$, to underline the behavior of each CBC during the whole computation as in Fig.~\ref{fig:NACA1_prms_CBC}. \newline

It can be noticed that the baseline LODI approach (a) allows to evacuate the initial acoustic wave and the convected vortices with minimal reflection. The only differences with the reference $p_{RMS}$ field are the small perturbations observed at the outlet and a background noise which can be due to the reflection of the initial acoustic wave. The local streamline LODI approach (b) is less efficient: the background noise is more pronounced and the dissymetry observed in the previous academic test cases is still present, which increases pressure fluctuations at the outlet. In the CBC-2D case (c), the map of pressure fluctuations seems more smooth and the vortices does not seem do produce spurious reflections. But the overall prms is increased in the whole domain, which can be due to a drift of the mean pressure because of the absence of relaxation in the boundary condition.\newline

\begin{figure}
\begin{center}

(a) Pressure fluctuation $p_{RMS}$\\
\includegraphics[width=0.37\textwidth]{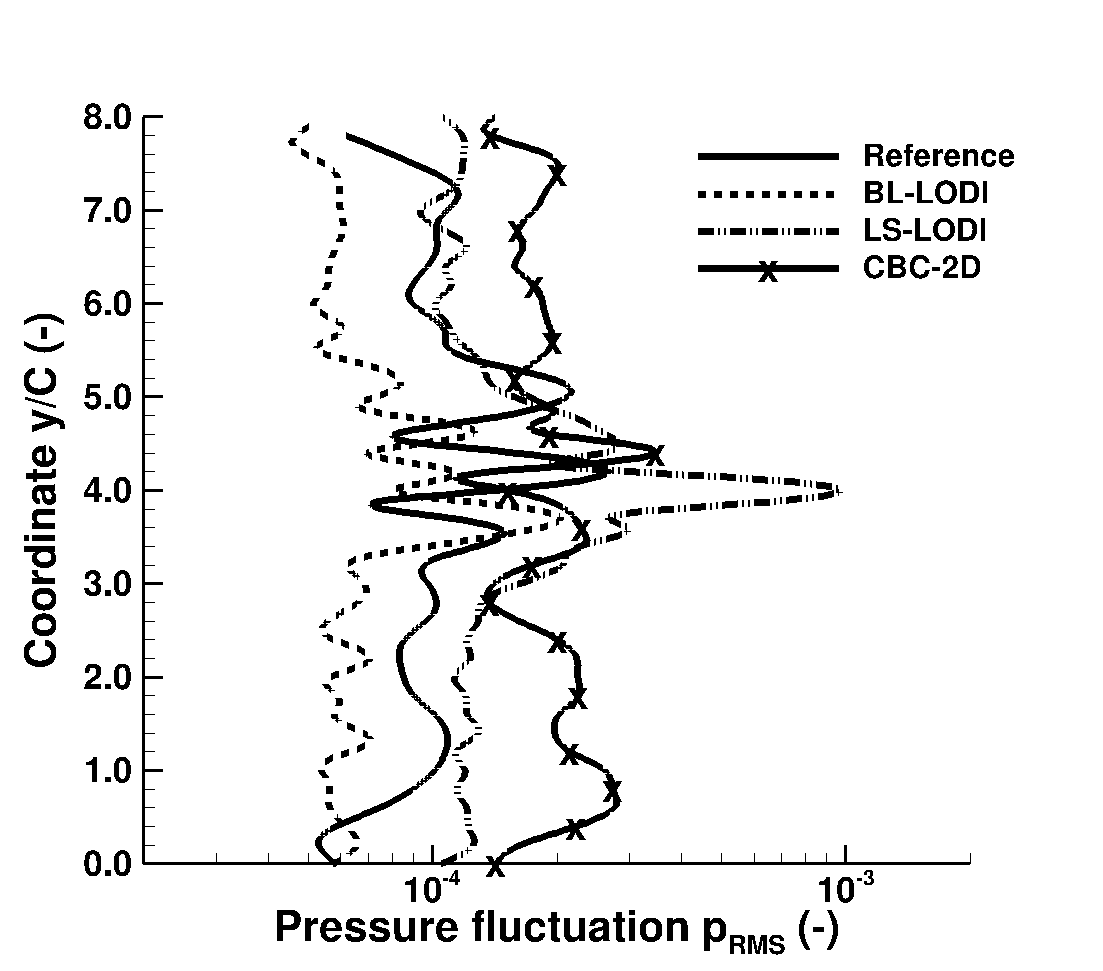}\\
(b) Axial velocity fluctuation $u_{RMS}$\\
\includegraphics[width=0.37\textwidth]{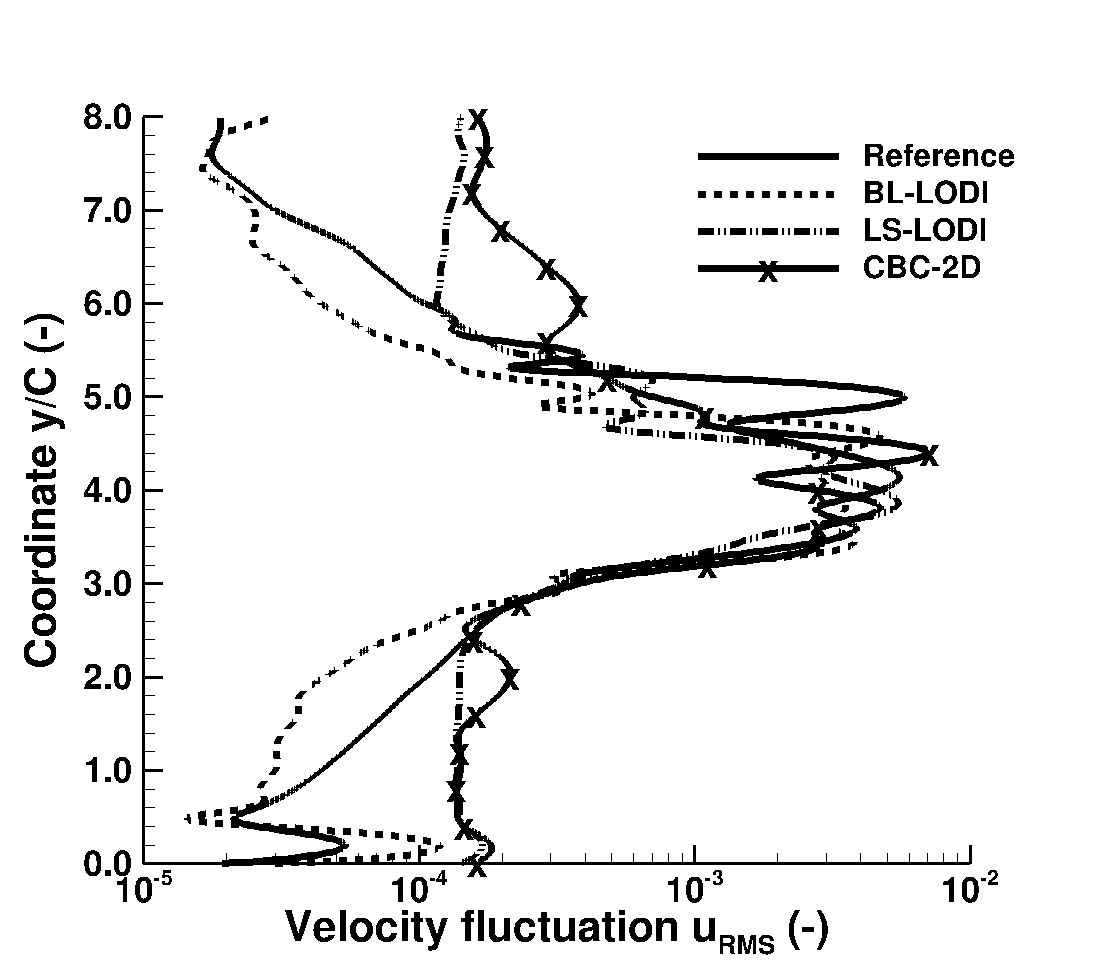}\\
(c) Transverse velocity fluctuation $v_{RMS}$\\
\includegraphics[width=0.37\textwidth]{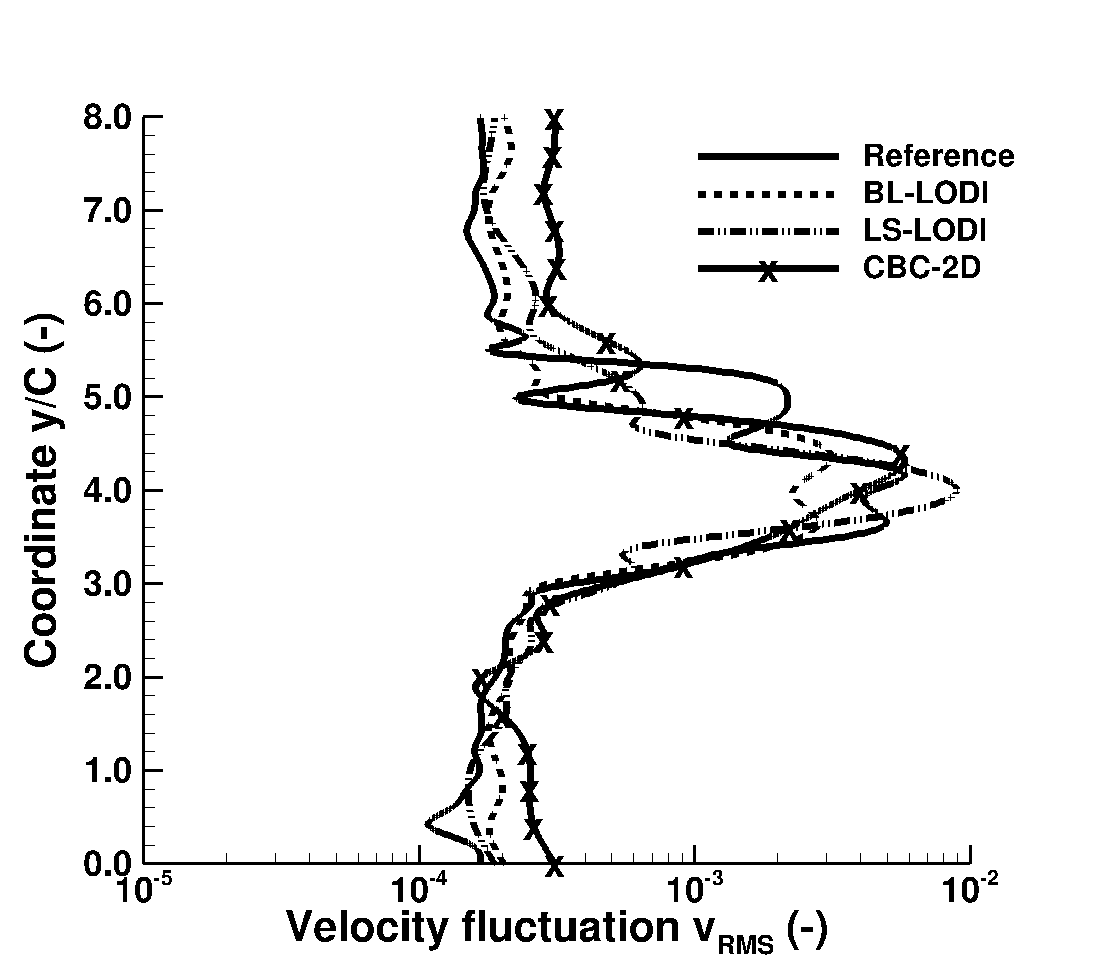}\\
\caption{Comparisons of the CBC approaches with the reference solution for the time-averaged fluctuations at $x=7.5C$: a) Pressure fluctuation $p_{RMS}$, b) Axial velocity fluctuation $u_{RMS}$ and c) Transverse velocity fluctuation $v_{RMS}$. \label{fig:NACA2}}
\end{center}
\end{figure}

To quantify more the performance of the three CBC approaches, the time-averaged fluctuations of pressure $p_{RMS}$ and velocity $u_{RMS}$, $v_{RMS}$ are shown in Fig.~\ref{fig:NACA2} at $x/C$=7.5 (close to the outlet where the CBC is applied). All CBCs show a good ability to predict the mean flow as well as pressure and velocity fluctuations. This test case remains a challenge for CBCs since the pressure and velocity fluctuations outside the wake are four magnitude orders lower than the mean field value. In that regard, all CBC methods predict pressure and velocity fluctuations of the same magnitude order than in the reference case. The CBC that provides the best results in term of accuracy for both pressure and velocity fluctuations is the BL-LODI approach. The LS-LODI gives satisfactory results outside the wake but it overestimates $p_{RMS}$ by a factor 2 in the wake (at $y/C$=4) because of the dissymetry in the vortices reflections. The CBC-2D approach overestimates $p_{RMS}$ outside the wake by a factor 4 but it gives good results in the wake, similar to those obtain with the BL-LODI approach. Similar conclusions can be drawn for axial and transverse velocity fluctuations, except that all methods predict the correct velocity fluctuations in the wake, including the LS-LODI approach.
 
\section{ Summary and conclusion}

An implementation of a non reflective outflow boundary condition, which does not need additional absorbing layers nor extended domains, has been proposed for a lattice Boltzmann solver. The methods presented here are based on the Characteristic Boundary Conditions (CBC) with the classical LODI approach (BL-LODI), its extension to transverse waves (CBC-2D) and the LODI approach in the local streamline based frame (LS-LODI). Three ways of computing missing populations in order to impose the CBC physical values have been introduced. The first one is based on the classical Zou/He boundary condition, while the other ones are based on the more stable regularized boundary condition : the so-called "Regularized BB" adaptation, where the off-equilibrium part of the stress tensor is evaluated thanks to a bounceback rule, and the "Regularized FD" adaptation, where it is computed thanks to an upwind finite difference scheme. All these methods provided very good results in the test case of a normal plane wave, where computed reflection rates were about $1\%$. Testcases of a plane wave with incidence and a spherical wave showed that the reflection rate increases with the incidence angle for the BL-LODI adaptation. When adding the effect of transverse terms, the reflected wave is considerably reduced but begins to increase for incidence angles greater than $30^\circ$. For these pure acoustic waves, the LS-LODI adaptation provided the best results since the reflection rate remained below $5\%$ whatever the incidence angle. However, this method is not adapted for a convected vortex, for which the CBC-2D adaptation with relaxation on density and transverse waves provided the best results and lead to only slight distortions of the velocity fields. As regards the numerical stability of the implemented CBC, the regularized adaptations have shown to be well more stable than the Zou/He one. The regularized FD adaptation associated with a regularized BGK scheme allowed to run the NACA0015 case at high Reynolds number ($Re=10^5$) in 2D with N=400 cells per chord length, which was not possible with Zou/He or Regularized BB adaptations. Thus, the use of the regularized FD adaptation is well advised for high Reynolds computations.

\section*{Acknowledgements}
The authors are grateful to the Calmip computing center of the Federal University of Toulouse (Project account number P1425) for providing all resources that have been used in this work. The authors would also thank J.F. Boussuge from CERFACS for his help on post-processing and the discussion about the method.

\label{}




\begin{thebibliography}{10}
\expandafter\ifx\csname url\endcsname\relax
  \def\url#1{\texttt{#1}}\fi
\expandafter\ifx\csname urlprefix\endcsname\relax\def\urlprefix{URL }\fi
\expandafter\ifx\csname href\endcsname\relax
  \def\href#1#2{#2} \def\path#1{#1}\fi

\bibitem{Tucker:2014}
P.~G. Tucker, J.~R. DeBonis, {Aerodynamics, computers and the environment},
  Philosophical Transactions of the Royal Society A: Mathematical, Physical and
  Engineering Sciences 372~(2022) (2014) 20130331--20130331.

\bibitem{Chen_AnnuRevFluid_30_1998}
S.~Chen, G.~D. Doolen, {L}attice {B}oltzmann method for fluid fows, Annual
  Review of Fluid Mechanics 30~(1) (1998) 329--364.

\bibitem{Succi_2001}
S.~Succi, The {L}attice {B}oltzmann {E}quation: {F}or {F}luid {D}ynamics and
  {B}eyond, Numerical Mathematics and Scientific Computation, Clarendon Press,
  2001.

\bibitem{Lallemand_PhysRevE_61_2000}
P.~Lallemand, L.-S. Luo, Theory of the lattice {B}oltzmann method:
  {D}ispersion, dissipation, isotropy, {G}alilean invariance, and stability,
  Phys. Rev. E 61 (2000) 6546--6562.

\bibitem{Buick_EPL_43_1998}
J.~M. Buick, C.~A. Greated, D.~M. Campbell, {L}attice {B}{G}{K} simulation of
  sound waves, EPL (Europhysics Letters) 43~(3) (1998) 235.

\bibitem{Marie_JCP_228_2009}
S.~Mari{\'e}, D.~Ricot, P.~Sagaut, Comparison between lattice {B}oltzmann
  method and {N}avier-{S}tokes high order schemes for computational
  aeroacoustics, Journal of Computational Physics 228~(4) (2009) 1056 -- 1070.

\bibitem{Heuveline_CMAP_58_2009}
V.~Heuveline, M.~J. Krause, J.~Latt, Towards a hybrid parallelization of
  lattice {B}oltzmann methods, Computers \& Mathematics with Applications
  58~(5) (2009) 1071 -- 1080, mesoscopic Methods in Engineering and Science.

\bibitem{Colonius_AnnuRevFluid_36_2004}
T.~Colonius, Modeling artificial boundary conditions for compressible flow,
  Annual Review of Fluid Mechanics 36~(1) (2004) 315--345.

\bibitem{Bodony_JCP_212_2006}
D.~J. Bodony, {A}nalysis of sponge zones for computational fluid mechanics,
  Journal of Computational Physics 212~(2) (2006) 681 -- 702.

\bibitem{Israeli_JCP_41_1981}
M.~Israeli, S.~A. Orszag, Approximation of radiation boundary conditions,
  Journal of Computational Physics 41~(1) (1981) 115 -- 135.

\bibitem{Poinsot_JCP_101_1992}
T.~Poinsot, S.~Lele, Boundary conditions for direct simulations of compressible
  viscous flows, Journal of Computational Physics 101~(1) (1992) 104 -- 129.

\bibitem{Yoo_CTM_9_2005}
C.~S. Yoo, Y.~Wang, A.~Trouv{\'e}, H.~G. Im, Characteristic boundary conditions
  for direct simulations of turbulent counterflow flames, Combustion Theory and
  Modelling 9~(4) (2005) 617--646.

\bibitem{Lodato_JCP_227_2008}
G.~Lodato, P.~Domingo, L.~Vervisch, {T}hree-dimensional boundary conditions for
  direct and large-eddy simulation of compressible viscous flows, Journal of
  Computational Physics 227~(10) (2008) 5105 -- 5143.

\bibitem{Izquierdo_PhysRevE_78_2008}
S.~Izquierdo, N.~Fueyo, Characteristic nonreflecting boundary conditions for
  open boundaries in lattice {B}oltzmann methods, Phys. Rev. E 78 (2008)
  046707.

\bibitem{Ginzburg:2008}
I.~Ginzburg, F.~Verhaeghe, D.~d'Humieres, Two-relaxation-time lattice boltzmann
  scheme: About parametrization, velocity, pressure and mixed boundary
  conditions, Comm. Comp. Phys. 3 (2008) 427--478.

\bibitem{Dhumieres:1992}
D.~D'Humieres, {Generalized lattice-Boltzmann equations}, Progress in
  Astronautics and Aeronautics~(159) (1992) 450--458.

\bibitem{Jung:2015}
N.~Jung, H.~W. Seo, C.~S. Yoo, {Two-dimensional characteristic boundary
  conditions for open boundaries in the lattice Boltzmann methods}, Journal of
  Computational Physics 302~(August) (2015) 191--199.

\bibitem{Heubes_JCAM_262_2014}
D.~Heubes, A.~Bartel, M.~Ehrhardt, {C}haracteristic boundary conditions in the
  lattice {B}oltzmann method for fluid and gas dynamics, Journal of
  Computational and Applied Mathematics 262 (2014) 51 -- 61, selected Papers
  from NUMDIFF-13.

\bibitem{Schlaffer:2013}
M.~B. Schlaffer, {Non-reflecting Boundary Conditions for the Lattice Boltzmann
  Method}, Ph.D. thesis, Technische Universit{\"{a}}t M{\"{u}}nschen (2013).

\bibitem{Zou_PhysFluids_9_1997}
Q.~Zou, X.~He, On pressure and velocity boundary conditions for the lattice
  {B}oltzmann {B}{G}{K} model, Physics of Fluids 9~(6) (1997) 1591--1598.

\bibitem{Latt_PhysRevE_77_2008}
J.~Latt, B.~Chopard, O.~Malaspinas, M.~Deville, A.~Michler, Straight velocity
  boundaries in the lattice {B}oltzmann method, Phys. Rev. E 77 (2008) 056703.

\bibitem{Latt:2006}
J.~Latt, B.~Chopard, {Lattice Boltzmann method with regularized pre-collision
  distribution functions}, Mathematics and Computers in Simulation 72~(2-6)
  (2006) 165--168.
\newblock \href {http://arxiv.org/abs/0506157} {\path{arXiv:0506157}}.

\bibitem{Malaspinas:2015}
O.~Malaspinas, {Increasing stability and accuracy of the lattice Boltzmann
  scheme: recursivity and regularization} (2015) 1--31\href
  {http://arxiv.org/abs/1505.06900} {\path{arXiv:1505.06900}}.

\bibitem{Bhatnaghar_PhysRev_94_1954}
P.~L. Bhatnagar, E.~P. Gross, M.~Krook, A {M}odel for {C}ollision {P}rocesses
  in {G}ases. {I}. {S}mall {A}mplitude {P}rocesses in {C}harged and {N}eutral
  {O}ne-{C}omponent {S}ystems, Phys. Rev. 94 (1954) 511--525.

\bibitem{Qian_EPL_17_1992}
Y.~H. Qian, D.~D'Humières, P.~Lallemand, Lattice {B}{G}{K} {M}odels for
  {N}avier-{S}tokes {E}quation, EPL (Europhysics Letters) 17~(6) (1992) 479.

\bibitem{Chapman_1952}
S.~Chapman, T.~Cowling, The mathematical theory of non-uniform gases: an
  account of the kinetic theory of viscosity, thermal conduction, and diffusion
  in gases, no. vol.~2, University Press, 1952.

\bibitem{Malaspinas_CompFluids_49_2011}
O.~Malaspinas, B.~Chopard, J.~Latt, General regularized boundary condition for
  multi-speed lattice {B}oltzmann models, Computers \& Fluids 49~(1) (2011) 29
  -- 35.

\bibitem{Yoo:2007}
C.~S. Yoo, H.~G. Im, {Characteristic boundary conditions for simulations of
  compressible reacting flows with multi-dimensional, viscous and reaction
  effects}, Combustion Theory and Modelling 11~(2) (2007) 259--286.

\bibitem{Albin_CompFluids_51_2011}
E.~Albin, Y.~D’Angelo, L.~Vervisch, Flow streamline based {N}avier-{S}tokes
  {C}haracteristic {B}oundary {C}onditions: {M}odeling for transverse and
  corner outflows, Computers \& Fluids 51~(1) (2011) 115 -- 126.

\bibitem{Philippi2006}
P.~C. Philippi, L.~A. Hegele, L.~O.~E. dos Santos, R.~Surmas, {From the
  continuous to the lattice Boltzmann equation: The discretization problem and
  thermal models}, Physical Review E 73~(5) (2006) 056702.

\bibitem{Lamb:1932}
H.~Lamb, {H}ydrodynamics, 6th Edition, Cambridge University Press, 1932.

\bibitem{Guo:2007}
Z.~Guo, C.~Zheng, B.~Shi, T.~S. Zhao, {Thermal lattice Boltzmann equation for
  low Mach number flows: Decoupling model}, Phys. Rev. E 75~(3) (2007) 1--15.

\bibitem{Yoo_CTM_11_2007}
C.~S. Yoo, H.~G. Im, Characteristic boundary conditions for simulations of
  compressible reacting flows with multi-dimensional, viscous and reaction
  effects, Combustion Theory and Modelling 11~(2) (2007) 259--286.

\end{thebibliography}


\bibliographystyle{elsarticle-num}

\end{document}